\documentclass[prb,twocolumn,superscriptaddress,nofootinbib]{revtex4}

\usepackage{graphicx}		% Include figure files
\usepackage{dcolumn}		% Align table columns on decimal point
\usepackage{bm}			% bold math
\usepackage{multirow}		%for tables
\usepackage{subfig}
\usepackage{amsmath}
\usepackage{array}
\begin{document}

\title{Long-time Behavior of Nuclear Spin Decays in Various Lattices}

\author{E. G. Sorte} \email{sorte@physics.utah.edu} \affiliation{Department of Physics, University of Utah, 115 South 1400 East, Salt Lake City, Utah 84112-0830, USA} 
\author{B .V. Fine} \affiliation{Institute for Theoretical Physics, University of Heidelberg, Philosophenweg 19, 69120 Heidelberg, Germany}
\author{B. Saam} \email{saam@physics.utah.edu} \affiliation{Department of Physics, University of Utah, 115 South 1400 East, Salt Lake City, Utah 84112-0830, USA} 

\date{\today}

\begin{abstract}

The transverse nuclear magnetic resonance (NMR) decays of $^{129}$Xe in polycrystalline xenon were recently shown to have a universal property:  in the long-time regime these decays all converge to the same sinusoidally modulated exponential function irrespective of the initial transverse spin configuration prepared by a sequence of one or more radio frequency pulses. The present work constitutes a more comprehensive survey of this phenomenon.  It examines transverse decays for several different isotopic concentrations of $^{129}$Xe, employs additional pulse sequences, and performs similar measurements in a different material: $^{19}$F in single-crystal and polycrystalline CaF$_2$. We additionally verified the polycrystalline nature of our frozen xenon samples by X-ray diffraction measurements.  With the possible exception of polycrystalline CaF$_2$ where the observation of the long-time behavior is limited by the experimental resolution, all these systems display the long-time universal behavior characterized by particular values of the exponential decay coefficient and beat frequency that were unique for each lattice. This behavior has been theoretically predicted based on the notion of microscopic chaos.  
 \end{abstract}

\maketitle

\section{\label{sec:intro} Introduction}

Nuclear magnetic resonance (NMR) has yielded valuable information to researchers for over five decades.  Motivations for manipulating the magnetic nuclear states of atoms range widely and cover a vast array of disciplines and applications.  In solid-state NMR, spin-spin relaxation is one of the important sources of microscopic information.  The disquieting fact is, however, that despite the ubiquitous applications of the methods and results of solid-state NMR, the accuracy of first-principle calculations of many spin-spin relaxation related quantities continues to be limited.~ \cite{Bloch, BPP, vleck, lowe_1, lowe_2, abragam, warren}  While many theories have been proposed to describe NMR spin-spin relaxation under various manipulations, none has succeeded in providing a controllable method for predicting measurements from a knowledge of the microscopic Hamiltonian of the many-spin system.  The properties of the spin-spin relaxation at times long compared to the characteristic spin-spin correlation time $T_2$ (the ``long-time" regime) are of particular interest in this respect because they are the most difficult to predict; these are the focus of the present work.

Recently, various authors have looked into the connections between many-body quantum dynamics and classical chaotic behavior.~\cite{Fine_1,prosen_2}  In particular, it was conjectured in Ref.\ 8 that microscopic chaos plays a role in the long-time behavior of nuclear transverse magnetization decays, and that, as a result, such decays all exhibit a generic behavior after a few times $T_2$, regardless of the initial state of the spins at the start of the decay.  This long-time behavior has the form:~\cite{Fine_1, bork}
\begin{eqnarray}
G(t) \simeq e^{-\gamma t} \cos(\omega t + \phi),
\label{eq:eq1}
\end{eqnarray}
where $G(t)$ denotes the nuclear magnetization transverse to the applied magnetic field, $\phi$ is a phase factor, $\gamma$ is the exponential decay coefficient and $\omega$ is the beat frequency.  Experimentally, this means that if the initial transverse spin configuration is varied through the application of one or more radio frequency (rf) pulses with varying phases and interpulse delay times, then the corresponding transverse decays (observed after the final pulse) will have different \emph{initial} behavior, but then \emph{will eventually all converge} to the form of Eq.~(\ref{eq:eq1}) with the same values of $\gamma$ and $\omega$.  The parameters $\gamma$ and $\omega$ have the following two important properties: 1) each is on the order of $T_2^{-1}$ and 2) each is universal in the sense that its respective value is independent of the initial transverse spin configuration generated by the rf excitation pulses.  These two properties, together with the fact that $T_2$ is the shortest intrinsic timescale of the system, imply that Eq.~(\ref{eq:eq1}) cannot be derived from conventional statistical-mechanical formulations, which require a separation of timescales that does not exist in these systems.  The possible role of microscopic chaos in the present context is discussed in Section~\ref{sec:discuss}.  

Morgan, et al.~\cite{morgan}\ were the first to look for these predicted effects with NMR in two solid polycrystalline xenon samples: one composed of $86\%$ $^{129}$Xe (isotopically enriched) and the other composed of 27\% $^{129}$Xe (naturally abundant).  They found that the transverse magnetization decays in these samples did indeed approach the universal form given by Eq.~(\ref{eq:eq1}) by measuring the FID and various solid echoes.      
 
The present work expands the work of Morgan et al.~\cite{morgan} in the following important ways:\\
1) by exploring six different xenon lattice systems obtained by systematically varying the ratios of xenon isotopes in the samples (see Table~\ref{table1}),\\
2) by exploring the behavior of $^{19}$F NMR in a different material (CaF$_2$) for several orientations of magnetic field,  \\
3) by employing a wide variety of pulse sequences in addition to the solid echo sequences used in Ref.\ 11 to generate distinct initial spin configurations in the sample, and \\
4) by significantly improving the quality of signals over those obtained in previous work.

\begin{table}
\caption{Description of the xenon lattices under consideration.  They are all face-centered-cubic but differ in the concentration of $^{129}$Xe.  NMR is performed exclusively on the $^{129}$Xe spins in the lattice.  The xenon isotopes with even numbers of nucleons (shown together in the final column) all have nuclear spin $I=0$.}
\begin{tabular}{c c c c c}
\hline
\hline
Name & System & $\%  ^{129}$Xe &  $\%  ^{131}$Xe &  $\%^{even}$Xe \\ \hline
43A & I & 85.6 & 1.9 & 12.5 \\ %\hline
%146C & II & 67 & 21 & 12 \\ %\hline
145C & II & 62.7 & 27.3 & 10.0 \\ %\hline
146A  & III & 54.7 & 36.1 & 9.2 \\ %\hline
146B/145A & IV & 46.5 & 45.2 & 8.3 \\ %\hline
145B & V & 29.6 & 64.0 & 6.4 \\ %\hline
123A & VI & 27.5 & 21.5 & 51.0 \\ %\hline
%\caption{Table caption}
\hline
\end{tabular}
\label{table1}
\end{table}

\section{\label{sec:theory} Theory}

An ensemble of thermally isolated interacting nuclear spins is described by the truncated magnetic dipole Hamiltonian in the rotating frame~\cite{slichter}:
\begin{eqnarray}
\mathcal{H}=\sum_{k<n}\left[J^\perp_{kn}(I^x_k I^x_n + I^y_k  I^y_n ) + J^z_{kn}I^z_k I^z_n \right], 
\label{eq:eq2}
\end{eqnarray}
where $I^\alpha_n$ is the quantum mechanical operator of the $n^{th}$ nuclear spin.  The coefficients for the interaction of the $k^{th}$ and $n^{th}$ spins are 
\begin{eqnarray}
J^\perp_{kn} =- \frac{1}{2}J^z_{kn} \simeq - \frac{(1-3\cos^2{\theta})}{2 | \textbf{r}_k - \textbf{r}_n|^3 },
\label{eq:eq3}
\end{eqnarray}
where $\theta$ is the angle between the $z$-axis defined by the external static magnetic field and the internuclear vector $\textbf{r}_k - \textbf{r}_n$.  We describe the transverse NMR decays by the function $G(t)$ which can be expressed as:

\begin{eqnarray}
G(t) = \text{Tr}\left\{\sum_n S_n^x \rho(t) \right\}
\label{eq:eq0}
\end{eqnarray}
where 

\begin{eqnarray}
\rho(t) = e^{i\mathcal{H}t}\rho_0e^{-i\mathcal{H}t}
\label{eq:rho}
\end{eqnarray}
and $\rho_0$ is the density matrix at time $t=0$.  Here we have chosen time origin ($t=0$) to be at the time corresponding to the end of the final pulse in each applied rf sequence.  Equation (\ref{eq:eq1}) is the long-time form of Eq. (\ref{eq:eq0}) predicted by Ref.~8.

To generate distinct initial density matrices for our spin ensembles, we used a variety of pulse sequences with varying interpulse delay times $\tau$.  We have analyzed the FID ($90^\circ_x$), various solid echoes~\cite{powles, slichter} ($90^\circ_{x} - \tau - 90^\circ_{y}$), the Jeener-Broekaert (JB) echo,~\cite{jeener, wang}($90^{\circ}_x - \tau_1 - 45^{\circ}_y - \tau_2 - 45^{\circ}_{\bar{y}}$ ), and the magic echo~\cite{rhim, slichter} ($90_y - \tau - 90_{\bar{y}}  - \tau'_{H1} - \tau'_{\bar{H1}} - 90_y$).  Here the subscripts for the angles indicate the axis of the rf pulse-induced rotation in the Larmor reference frame.  $\bar{y}$ indicates the negative direction along the y-axis.  The meaning of $H_1$ is explained at the end of Section \ref{sec:theory}. The solid echo, JB echo, and magic echo manipulate spin coherences generated by spin-spin interactions and, in this sense, are not like Hahn echoes that simply compensate for an inhomogeneous external field.  Below we will briefly describe each sequence; they are reviewed and analyzed in great detail elsewhere.~\cite{slichter}  

In the solid echo sequence ($90^\circ_{x} - \tau - 90^\circ_{y}$), two $90^\circ$ pulses are applied out of phase with an interpulse delay time $\tau$ that can be varied.~\cite{powles, slichter}  The density matrix which follows the second pulse is given by:

\begin{eqnarray}
 \rho_0 = e^{ i\mathcal{H}_R\tau} \exp\left\{\frac{\sum_n I_n^x H}{k_B T} \right\}e^{ -i\mathcal{H}_R\tau} 
   \label{eq:se12}
\end{eqnarray}
where the ``rotated Hamiltonian'' $\mathcal{H}_R$ is given by:~\cite{support}

\begin{eqnarray}
\mathcal{H}_R=\sum_{k<n}\left[J^\perp_{kn}I^x_k I^x_n + J^z_{kn}I^y_k  I^y_n + J^\perp_{kn}I^z_k I^z_n \right]
   \label{eq:se13}
\end{eqnarray}

For pairs of interacting spins 1/2, the solid echo yields a perfect recovery of the initial magnetization at time $\tau$ after the second pulse, similar to the conventional Hahn echo.  However, in many-spin systems complete refocusing does not occur, and the deviations from complete refocusing become more pronounced as $\tau$ is lengthened.~\cite{Cho}  As follows from Eq.~(\ref{eq:se12}), the solid echo pulse sequence creates new initial spin density matrices as $\tau$ is varied.~\cite{support}  

The JB echo ($90^{\circ}_x - \tau_1 - 45^{\circ}_y - \tau_2 - 45^{\circ}_{\bar{y}}$ ) partially transforms Zeeman order into dipolar order.~\cite{jeener, wang}  The $45_y^{\circ}$ pulse performs the partial transformation into dipolar order --- a state similar to that produced by adiabatic demagnetization in the rotating frame.  The $45_{\bar{y}}^{\circ}$ pulse generates a JB ``echo" from the dipolar order, better regarded as an FID different from the usual one generated from Zeeman order.  Here we do not explicitly include the expression for the density matrix after the final pulse. 

The magic echo sequence is $90_y - \tau - 90_{\bar{y}}  - \tau'_{H1} - \tau'_{\bar{H1}} - 90_y$.  $H_1$ is the amplitude of the radio frequency pulse applied at resonance, and $\tau'_{H_1}$ represents the duration of time that the pulse is applied.  $\bar{H_1}$ represents the amplitude of the resonant pulse applied with a $180^{\circ}$ phase shift from $H_1$.  We used the magic echo as a check on our control of spin manipulation; the sequence is designed to reverse the Hamiltonian in Eq.~(\ref{eq:eq2}) and reproduce the entire FID shape.  The magic echo is an exceptional pulse sequence that completely reverses the spin dynamics in the limit of very strong $H_1$, and therefore the onset of the universal asymptotic behavior occurs not after a time on the order of several $T_2$, but rather after a time $t'=\tau'_{H_1} - \tau$ (peak of the echo) plus several times $T_2$.~\cite{slichter}  

\section{\label{sec:exp} Experimental Procedures}

We explored a wide parameter space in search of experimental evidence for the predicted universal long-time behavior in isolated quantum systems.  First, we looked at six different isotopic compositions of solid hyperpolarized xenon using the solid echo sequence with varying values of $\tau$.  Then, using one of these compositions (System I), we applied three different pulse sequences (FID, solid echo, JB echo) to observe the long-time behavior associated with each initial transverse spin configuration.  Next, we examined $^{19}$F in CaF$_2$ single-crystals at three different orientations of applied magnetic field as well as in polycrystalline CaF$_2$ powder.  We used three different pulse sequences to observe their long-time behavior (FID, solid echo, magic echo).  

\subsection{\label{sec:xe} Solid Xenon}

Hyperpolarized $^{129}$Xe is an ideal system for investigating the hypothesis of a universal long-time behavior.  In this van der Waals solid at 77 K, the xenon spin system is thermally well-isolated from its environment ($T_1 \gg T_2$).  The NMR signals from a macroscopic sample are large (greater than 1mV before amplification); the long-time reqime can thus be examined with good signal-to-noise-ratio (SNR).  Additionally, $T_2$ is long ($\approx 0.5$~ms), allowing for relatively simple radio-frequency manipulation of the sample, as the corresponding bandwidths are small and the deadtimes short in comparison.  These conditions make the use of short ``hard" rf pulses practical.  Finally the $^{129}$Xe nuclei have spin = 1/2, which makes them maximally non-classical in terms of spin angular momentum quantization, and are therefore ideal for exploring quantum connections to conventionally classical phenomena such as chaos.  The magnitude of the spin quantum number also means we did not have to account for a nuclear quadrupole moment.

We used dynamic nuclear polarization in the form of spin-exchange optical pumping~\cite{rmp} to generate nuclear polarizations of $\approx$$10\%$ in solid $^{129}$Xe nuclei.  We then measured the transverse relaxation of these nuclei over many orders of magnitude (see Fig.~\ref{fig:43A_FID}).  These experiments generate extremely large signals that can be precisely measured over unconventionally long times, i.e., many times $T_2$.  Spin-exchange convection cells~\cite{cell_paper} allowed for rapid and repetitive generation of a few millimoles of highly polarized liquid xenon, which we then froze rapidly by exposing the sample to a stream of liquid nitrogen.  During rapid freezing, the majority of the polarization in the liquid xenon sample survives the phase transition to the solid.~\cite{support}

The solid xenon samples were generated in an applied magnetic field of 2 T ($^{129}$Xe Larmor frequency 24.56 MHz) and maintained at a temperature of 77 K.  Induction signals were acquired with an Apollo (Tecmag) NMR spectrometer using $10$ $\mu$s square excitation pulses with a receiver dead time of 60 $\mu$s.  At 77 K, $^{129}$Xe $T_1 \approx 2.5$ hrs~\cite{gatzke} while $T_2 \approx 0.5$ ms~\cite{yen}, justifying our assumption of thermal isolation during the transverse decay.  Our samples consisted of 1 to 2 mmol xenon in various isotopic ratios (see Table~\ref{table1}).  We acquired the transverse decay either after one $90^\circ$ pulse (in the case of the FID) or after the last of a series of pulses that create various echoes.  The enormous signals thus acquired were too large to fit into the 16-bit digitizer native to the Apollo system.  Acquisition was therefore carried out in a two separate experiments with two different NMR-gain settings.  There was sufficient overlap of the temporal signal in both experiments to ensure that the reconstruction of the entire decay was unambiguous.  

\begin{figure}[htbp]
\begin{center}
\includegraphics[width=0.47\textwidth]{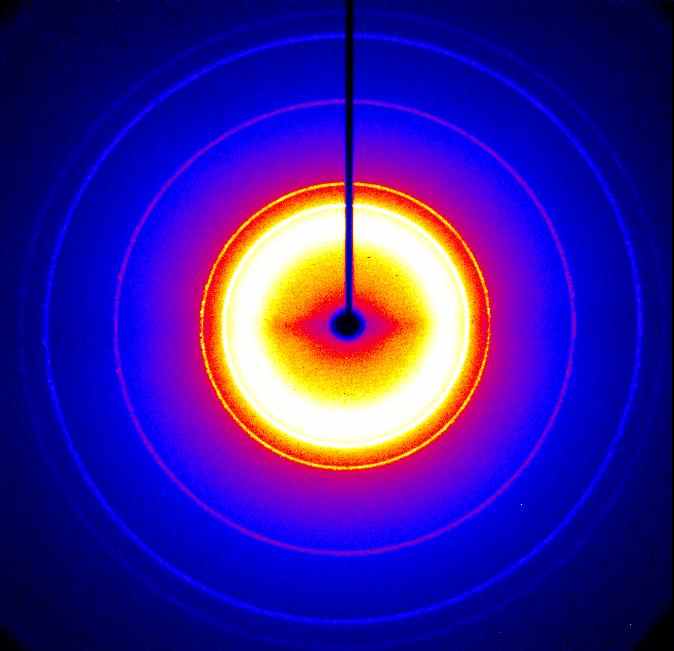}
\end{center}
\caption{(Color online) Result of X-ray diffraction experiment of rapidly frozen xenon.  Sharp rings indicate a polycrystalline sample with well-defined lattice spacing, determined to be 4.4~\AA\ (nearest-neighbor spacing) in agreement with the known value.~\cite{fitz}}
\label{fig:xray}
\end{figure}

To investigate the atomic arrangement of the xenon in our samples (see discussion in Section \ref{sec:results} below), we undertook X-ray diffraction studies on a natural xenon sample with the isotopic makeup of System VI.  A room-temperature sample was exposed to nitrogen gas at $100$ K (well below the 160 K freezing point of xenon), freezing it suddenly in a manner similar to our NMR experimental procedure.  X-ray diffraction was performed with a 4.1 kW Bruker Copper K-Alpha X-ray source operating at a wavelength of $\lambda = 1.5814$~\AA.  The sample was maintained at 100 K by a built-in cryostat.  Typical results from one of ten runs are shown in Fig.~\ref{fig:xray}. The sharp well-defined rings are characteristic of a polycrystalline solid; i.e., randomly oriented crystals of definite atomic spacing. Analysis of the rings due to xenon (after background subtraction) yields the known nearest-neighbor spacing of $\sim 4.4$~\AA~\cite{fitz} for xenon in a face-centered-cubic (fcc) lattice.  

\subsection{\label{sec:caf2} Calcium Fluoride}

The CaF$_2$ crystals used in our experiments were obtained from Optovac, Inc (now owned by Corning) and are doped with 0.1\% yttrium or gadolinium to reduce $T_1$ to $\approx$~1 s.  These are nominally cylindrical samples with the cylinder axis perpendicular to the [100], [110], or [111] crystal direction; these respective directions would thus be parallel to the applied magnetic field in a typical transverse solenoidal NMR coil.  Because these samples were altered somewhat to fit into our NMR probe, the crystal directions are no longer exact (though care was taken to maintain them as much as possible).  $T_2 \approx 25~\mu$s for all CaF$_2$ samples.  The signals are acquired at 2~T ($^{19}$F Larmor frequency 83.55 MHz) with the same Apollo (Tecmag) NMR spectrometer using $2~\mu$s square pulses with a receiver dead time of $15~\mu$s at room temperature.  The CaF$_2$ powder was obtained from J.\ T.\ Baker Chemical Co.\ and contains nominally 98\% CaF$_2$ with traces of chlorides, sulfates, heavy metals, and iron.  $T_1$ of the powder was measured by saturation-recovery to be $\approx$ 1s. 
  
 \section{\label{sec:results} Experimental results and discussion}
 
 \subsection{\label{sec:fid} Solid Hyperpolarized $^{129}$Xe} 
    
\subsubsection{\label{sec:fid} The Free Induction Decay} 
\begin{figure}[htbp]
\begin{center}
\includegraphics[width=0.47\textwidth]{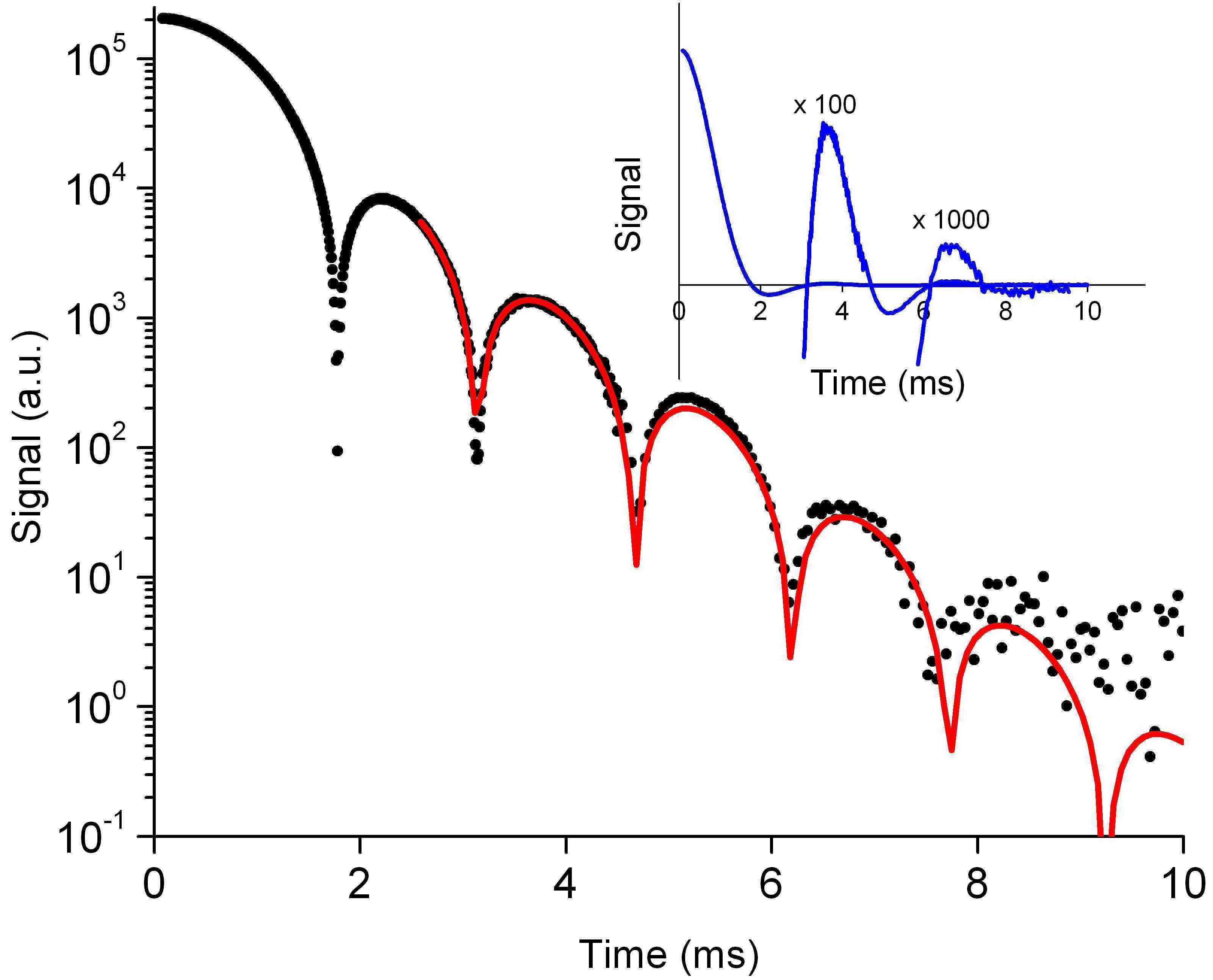}
\end{center}
\caption{ (Color online) $^{129}$Xe FID corresponding to System I ($86\%\ ^{129}$Xe; see Table~\ref{table1}) shown on a semi-log plot.  The solid line is a fit of the absolute value of Eq.~(\ref{eq:eq1}) to the long-time signal at $t > 2.5$~ms.  The inset shows the FID absorption signal on a linear scale, illustrating the zero-crossings of the FID.  Later portions of the linear signal have been enhanced as indicated for clarity.}
\label{fig:43A_FID}
\end{figure}

In Fig.~\ref{fig:43A_FID} we show a typical $^{129}$Xe FID at 2 T and 77 K.  The magnitude of the quadrature detected signal is plotted vs.\ time on a semi-log plot to illustrate the decay over many orders of magnitude.  The SNR here is about an order of magnitude better than the previous study.~\cite{morgan}  Since the plot shows the log of the absolute value of the FID signal, the cusps indicate the zero-crossings of the FID (see the inset graph of Fig.~\ref{fig:43A_FID} for a linear-scale plot of the absorption signal explicitly showing the zero-crossings).  The decay coefficient $\gamma = 1.25 \pm$ 0.04~ms$^{-1}$ and the oscillation frequency $\omega = 2.06 \pm 0.04$~rad/ms were extracted from a fit of the long-time portion of the FID to Eq.~(\ref{eq:eq1}).    These parameters are on the order of $T_2^{-1}$ and agree well with the values obtained by Morgan, et. al.~\cite{morgan}  $T_2$ was estimated as $\sqrt{1/M_2}$ where $M_2$ is the second moment of the lineshape given by~\cite{abragam}

 \begin{eqnarray}
M_2 = -\frac{1}{G(0) }\frac{d^2G(t)}{dt^2}\bigg|_{y=0} 
   \label{eq:moment}
\end{eqnarray}

 \subsubsection{\label{sec:solid_echo} The Solid Echo}
\begin{figure*}[!]
 \subfloat[]
  {\label{fig:one}\includegraphics[width=0.47  \textwidth]{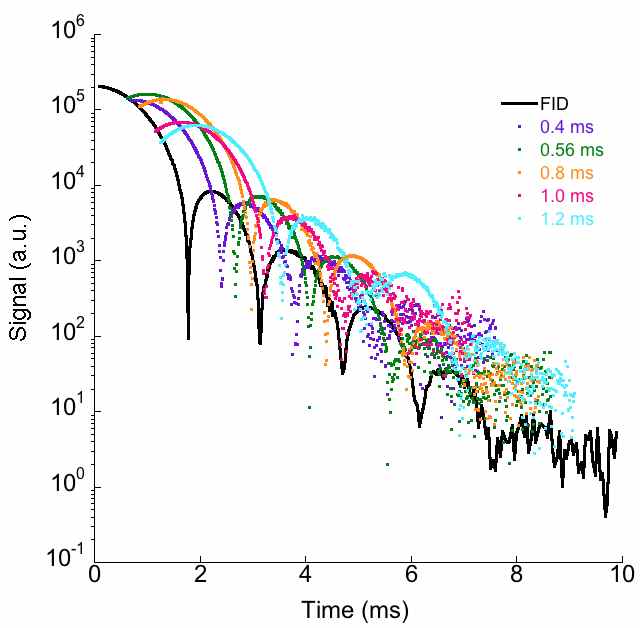}}                
\subfloat[]
  {\label{fig:two}\includegraphics[width=0.47\textwidth]{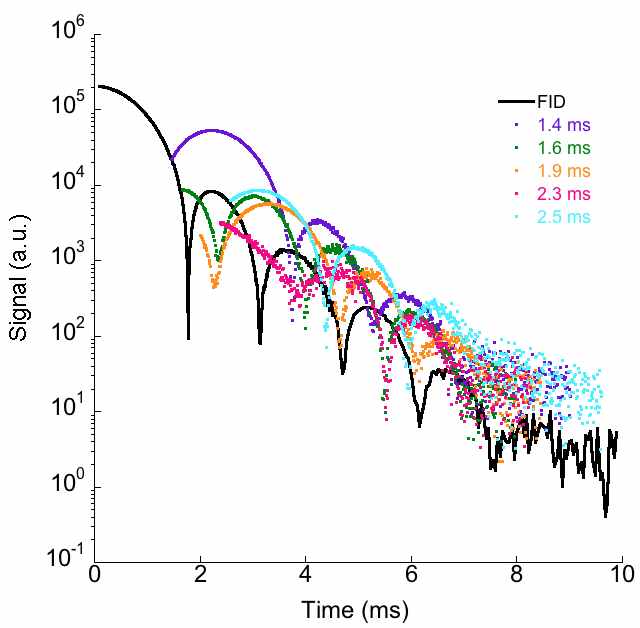}}\\
     \subfloat[]
  {\label{fig:three}\includegraphics[width=0.47  \textwidth]{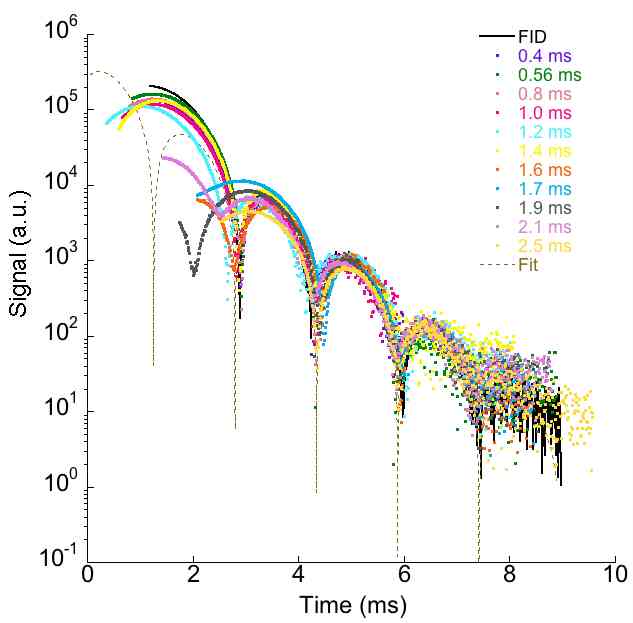}}   
  \caption{(Color online) $^{129}$Xe solid echoes in xenon enriched to $86\%$ $^{129}$Xe (System I).  In (a) and (b) we show ten solid echoes on a semilog plot together with the FID of Fig.~\ref{fig:43A_FID}.  Signals are split between (a) and (b) for visual clarity only.  In (c), we show the data in both (a) and (b) together with the echoes time-shifted to illustrate the convergence of the long-time behavior.  The broken line is the fit to the FID, also shown in Fig.~\ref{fig:43A_FID}, demonstrating that the solid echoes all exhibit the same long-time behavior as the FID with the same values of $\gamma$ and $\omega$ (see Table~\ref{table2}).  }
  \label{fig:43_a}
\end{figure*}

In Fig.~\ref{fig:43_a} we show signals associated with ten solid echoes obtained for different values of the interpulse delay time $\tau$ together with the free induction decay.  These data were collected for $86\%$ $^{129}$Xe (System I).  The initial portion of the FID following the first pulse was acquired along with the solid echo following the second pulse in order to provide a relative amplitude normalization and thus to correct for polarization levels that varied somewhat from experiment to experiment.  Care was taken to eliminate residual longitudinal magnetization from the echo signal, and this was checked by using a $90^\circ_x - \tau - 90^\circ_x$ sequence prior to each echo experiment and verifying a null signal (i.e., precise 90$^\circ$ pulses).  We note that some of the data reported in Morgan, et al.~\cite{morgan}\ from a similar experiment likely are corrupted by a small component of magnetization left along the z-axis by the first pulse that is then excited into the transverse plane by the second pulse.  The observed $\pi$ phase shift of the echoes with respect to the FID shown in that work was therefore likely coincidental.  This does not affect their conclusions concerning the universal long-time behavior of the signals, but it is critical for exploring relations concerning the amplitudes and phases of solid echo signals relative to the FID in the long-time regime.~\cite{unpub_1}

In Fig.~\ref{fig:three}, we make explicit time shifts to the data shown in Fig.~\ref{fig:one} and Fig.~\ref{fig:two} to compare the behavior of each signal in the long-time regime.  The dashed line in Fig.~\ref{fig:three} is a plot of Eq.~(\ref{eq:eq1}) with the parameters obtained from the fit in Fig.~\ref{fig:43A_FID}.  These values are in good agreement with those obtained by Morgan, et al.~\cite{morgan}\ on the same system.  Similar long-time fits to the individual solid echoes produced values for $\gamma$ and $\omega$ consistent with those obtained for the FID; it is in this sense that the long-time behavior is universal, in contrast to the varied decay shapes exhibited in the initial portions of the decays.   

Table~\ref{table2} lists the values of the decay constant $\gamma$ and the beat frequency $\omega$ for six different xenon isotopic compositions.  For each composition, the parameters were obtained from a fit to the FID.  As a check, fits were also performed to each individual echo to verify that the extracted parameter values were in agreement with those obtained from the fit to the FID.  

\begin{figure*}[htbp!]
 % \centering
  \subfloat[]
  {\label{fig:145_a}\includegraphics[width=0.48\textwidth]{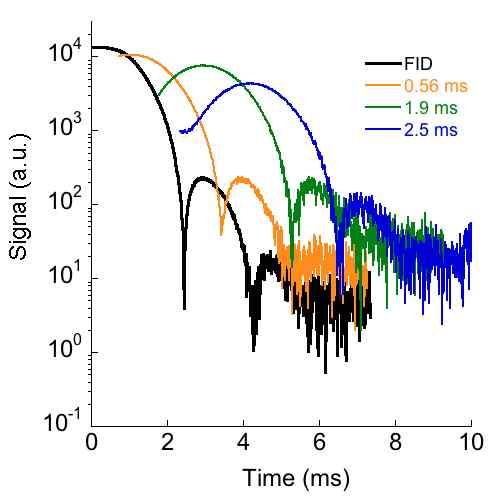}}                
\subfloat[]
  {\label{fig:145_b}\includegraphics[width=0.50\textwidth]{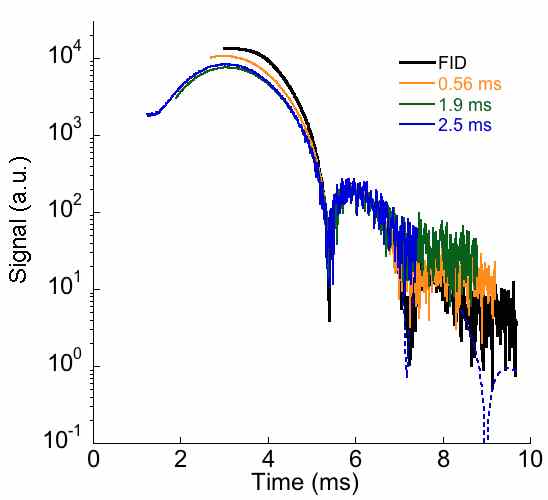}}
   %  \subfloat[]     
 \caption{(Color online) $^{129}$Xe free-induction decay and solid echoes in xenon having $29.6\%$ $^{129}$Xe (System V).  Due to lower concentrations of $^{129}$Xe in this system, the SNR is reduced compared to System I.  (a) Solid echo signals shown on a semilog plot with the FID.  (b)  The same data as in (a) are shown again with the echoes time-shifted to illustrate the convergence of the long-time behavior.  The broken line is a fit to the long-time portion of the FID from which $\gamma$ and $\omega$ have been determined (see Table~\ref{table2}).  }
  \label{fig:145B}
\end{figure*}

In Fig.~\ref{fig:145B} we show the data for System V, representative in SNR of the systems with lower concentrations of $^{129}$Xe.  Though the lower isotopic concentration of $^{129}$Xe has reduced the SNR relative to System I, we still observe convergence of different transverse decays to the respective universal values of $\gamma$ and $\omega$ for each system.  The echo amplitudes are normalized in the same way as for the other xenon compositions as described above.  The data similar to Fig.~\ref{fig:43_a} and Fig.~\ref{fig:145B} which pertain to the other xenon isotopic concentrations are included in the supplemental section to this work.~\cite{supp}

%\vspace{5mm}
\begin{table}[htbp]
\caption{Values of the decay coefficient $\gamma$ and beat frequency $\omega$ for each xenon isotopic composition tested in our experiments.  The fourth column refers to the percentage of even (spin-zero) xenon isotopes.  Data for the compositions marked with $^*$ are found in the supplemental section [Ref. 24].}
\begin{tabular}{c c c c c c }
%\hline
%\multicolumn{7}{|c|}{Spin Systems} \\
\hline\hline
System & $\%  ^{129}$Xe &  $\%  ^{131}$Xe &  $\%^{\rm{even}}$Xe & $\gamma$ (ms$^{-1}$) & $\omega$ (rad/ms) \\ \hline
I & 85.6 & 1.9 & 12.5  & 1.25 $\pm$ 0.04 & 2.06 $\pm$ 0.04 \\ %\hline
%II & 67 & 21 & 12 & ? & ? \\ \hline
II & 62.7 & 27.3 & 10.0 & 1.34 $\pm$ 0.34 & 1.95 $\pm$ 0.14 $^{*}$ \\ %\hline
III & 54.7 & 36.1 & 9.2 & 1.42 $\pm$ 0.19 & 1.96 $\pm$ 0.05 $^{*}$ \\ %\hline
IV & 46.5 & 45.2 & 8.3 & 1.24 $\pm$ 0.05 & 1.71 $\pm$ 0.10 $^{*}$\\ %\hline
V & 29.6 & 64.0 & 6.4 & 1.50 $\pm$ 0.06 & 1.74 $\pm$ 0.05 \\ %\hline
VI & 27.5 & 21.5 & 51.0 & 1.03 $\pm$ 0.10 & 1.53 $\pm$ 0.05 $^{*}$ \\ %\hline
\hline
\end{tabular}
\label{table2}
\end{table}

\subsubsection{\label{sec:jb} The Jeener-Broekaert Echo}

\begin{figure}[htbp!]
\begin{center}
\includegraphics[width=0.47\textwidth]{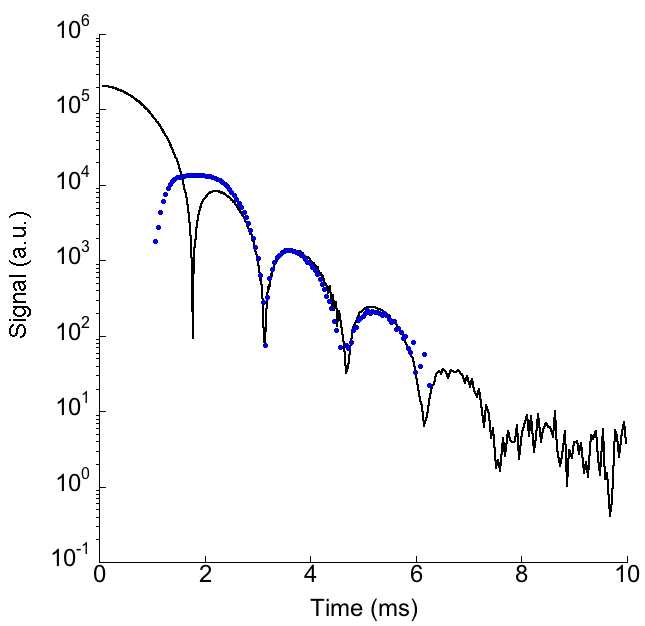}
\caption{(Color online) $^{129}$Xe FID (solid black line) and Jeener-Broekaert echo (blue points) in solid xenon enriched to $85.6\%$ $^{129}$Xe (System I).  The echo is scaled and time-shifted to show the uniformity of the long-time behavior.  The interpulse delays are $\tau_1= 0.5$~ms and $\tau_2 = 3$~ms.}
\label{fig:jb_echo_FID}
\end{center}
\end{figure}

Figure~\ref{fig:jb_echo_FID} shows the JB echo ($90^{\circ}_x - \tau_1 - 45^{\circ}_y - \tau_2 - 45^{\circ}_{\bar{y}}$ ) for System I along with the corresponding FID from Fig.~\ref{fig:43A_FID}.    The interpulse delays are $\tau_1=0.5$~ms and $\tau_2 = 3$~ms.  Usually, $\tau_2$ is much longer than $T_2$ in the JB echo sequence.  In our experiments, $\tau_2$ $\approx 6 \times T_2 $ --- sufficiently long to generate a spin state distinct from that generated by the FID or solid echoes.  While the initial portion of the JB echo is distinct from that of both the FID and the various solid echoes, its long-time behavior exhibits the form of Eq. (\ref{eq:eq1}) and with the same values of $\gamma$ and $\omega$ (see Fig.\ {\ref{fig:jb_echo_FID}}).

\subsection{\label{sec:caf2} Calcium Fluoride}

\subsubsection{\label{sec:fid} The Free Induction Decay and Solid Echoes} 

\begin{figure}[htbp]
\begin{center}
\includegraphics[width=0.47\textwidth]{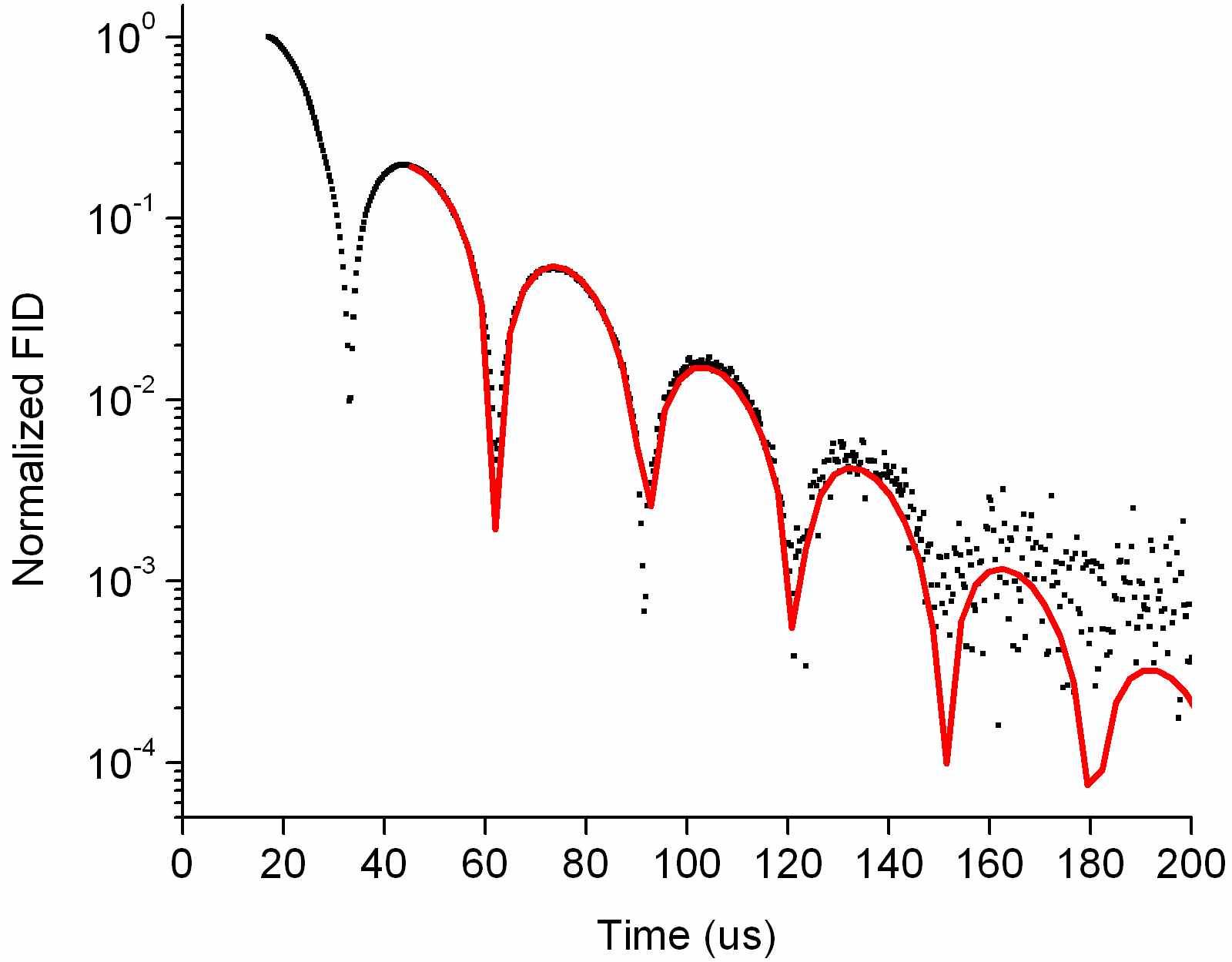}
\caption{Normalized $^{19}$F FID in CaF$_2$ corresponding to the applied field $\textbf{H}_0$ along the [110] direction (System VIII; see Table~\ref{table4}).  The solid line is a fit of the absolute value of Eq.~(\ref{eq:eq1}) to the long-time signal at $t > 45$ $\mu$s.  $T_2$ is approx $25$ $\mu$s in this system.}
\label{fig:caf2_110_fid}
\end{center}
\end{figure}

In single-crystal CaF$_2$, different orientations of the applied magnetic field $H_0$ with respect to the crystal planes imply different interaction coefficients in the interaction Hamiltonian $\mathcal{H}$ [Eq. (\ref{eq:eq2})].  Hence, experiments performed with each orientation of $H_0$ amount to independent checks of the validity of the long-time equation Eq.~(\ref{eq:eq1}).  In Fig.~\ref{fig:caf2_110_fid} we show a typical $^{19}$F FID at 2 T and 290 K. As with xenon, a semilog plot is used and the cusps indicate zero-crossings.  Solid echo data for single-crystal [111] CaF$_2$ (System IX; see Table~\ref{table4}) are shown in Fig.~\ref{fig:caf2_111_se}.  These data are representative in behavior and SNR of Systems VII - IX.  The beat frequencies in each crystal orientation were first measured by Lowe and Norberg~\cite{lowe_1} and then more accurately by Engelsberg and Lowe.~\cite{lowe_2}  Figure~\ref{fig:caf2_111_se} shows that the long-time behavior is universal according to Eq.~(\ref{eq:eq1}) for $^{19}$F in a single-crystal sample of CaF$_2$.  The same conclusion was drawn from the data for the other two crystal orientations.~\cite{supp}  The beat frequency $\omega$ measured by us for each orientation of $H_0$ is compared to that obtained by Engelsberg and Lowe~\cite{lowe_2} in Table~\ref{table4}.  The small discrepancies between the values obtained by us and those obtained by Engelsberg and Lowe are likely due to slight misalignments of the samples in our experiment; see Figs. S5c, S6c, and S7d in the supplemental section (Ref.~24) for a graphical comparison of our data with those of Engelsberg and Lowe.  
  
\begin{table*}[htbp]
\centering
\caption{Values of the decay coefficient $\gamma$ and beat frequency $\omega$ for each CaF$_2$ system tested.  Columns 4 and 5 are our experimental data for the decays.  In the last column we list the values of $\omega$ calculated from the data obtained by Engelsberg and Lowe (taken from Ref.~7) for purposes of comparison.  The impurities for the powder sample (System X) are as follows: 0.9\% Cl, 0.5\% SO$_4$, 0.3\% heavy metals (such as Pb), 0.3\% Fe.  Data for the crystals marked with $^*$ are found in the supplemental section [Ref. 24].  }
\begin{tabular}{c c c c c c }
\hline \hline
System & Direction & Impurtiy & $\gamma$ (ms$^{-1}$) & $\omega$ (rad/ms) & $\omega_{EL}$ (rad/ms) \\ [0.5ex] 
 \hline
VII & [100] & 0.01 \% Gd & 51.0 $\pm$ 0.1 & 144.4 $\pm$ 0.1 & 153.32 $\pm$ 0.09 $^{*}$ \\ %\hline
VIII & [110] &  0.01 \% Y & 43.3 $\pm$ 0.1 & 106.2 $\pm$ 0.1 & 101.9 $\pm$ 0.7 $^{*}$ \\ %\hline
IX & [111] & 0.01 \% Gd & 31.9 $\pm$ 0.1 & 66.2 $\pm$ 0.1 & 65.6 $\pm$ 0.3 $^{*}$\\  %hline
X & powder & 2\% & n/a & n/a & n/a\\ \hline
%\hline
\end{tabular}
\label{table4}
\end{table*}

\begin{figure*}[htbp]
 \subfloat[]
  {\label{fig:a}\includegraphics[width=0.47  \textwidth]{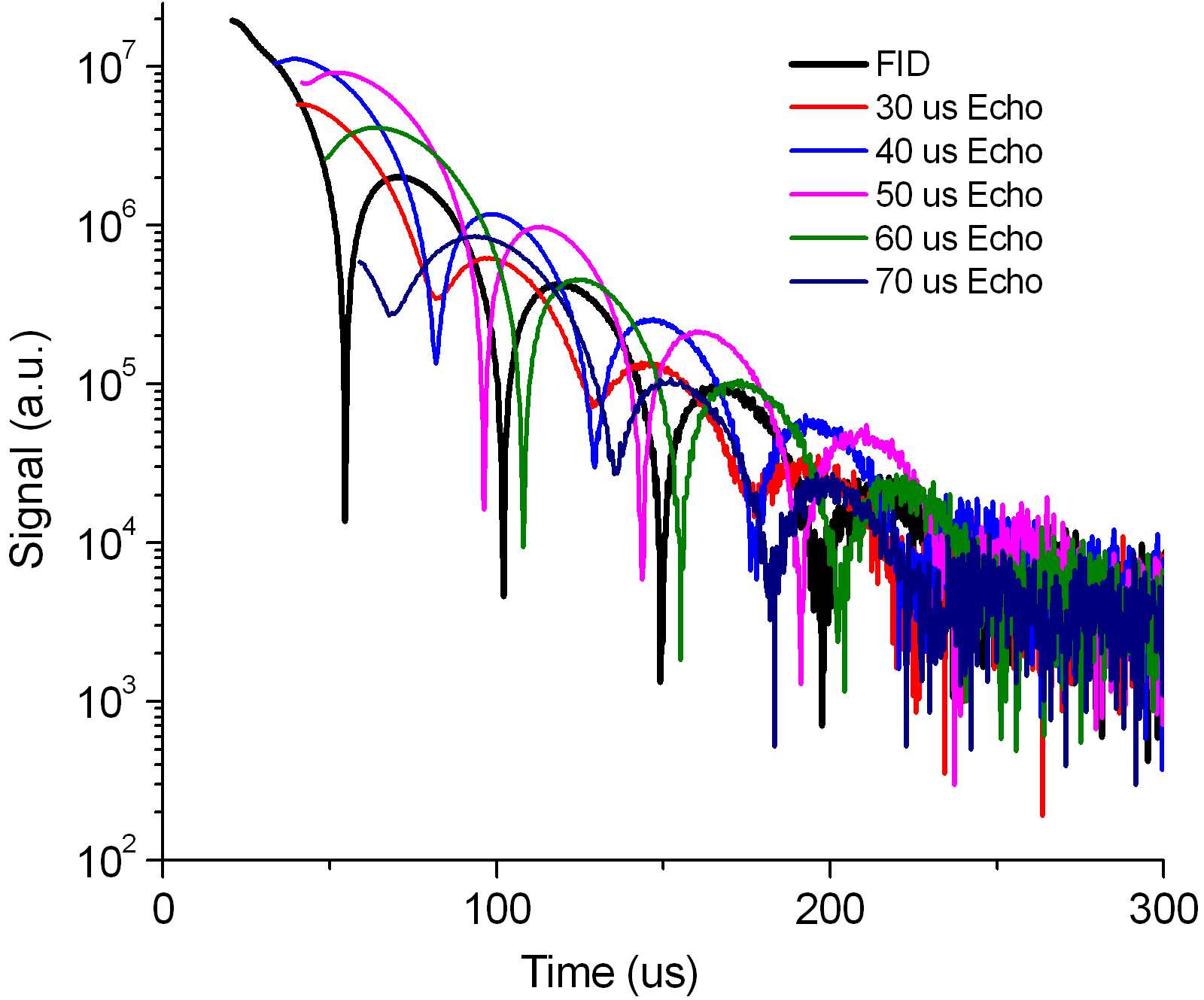}}                
\subfloat[]
  {\label{fig:b}\includegraphics[width=0.47\textwidth]{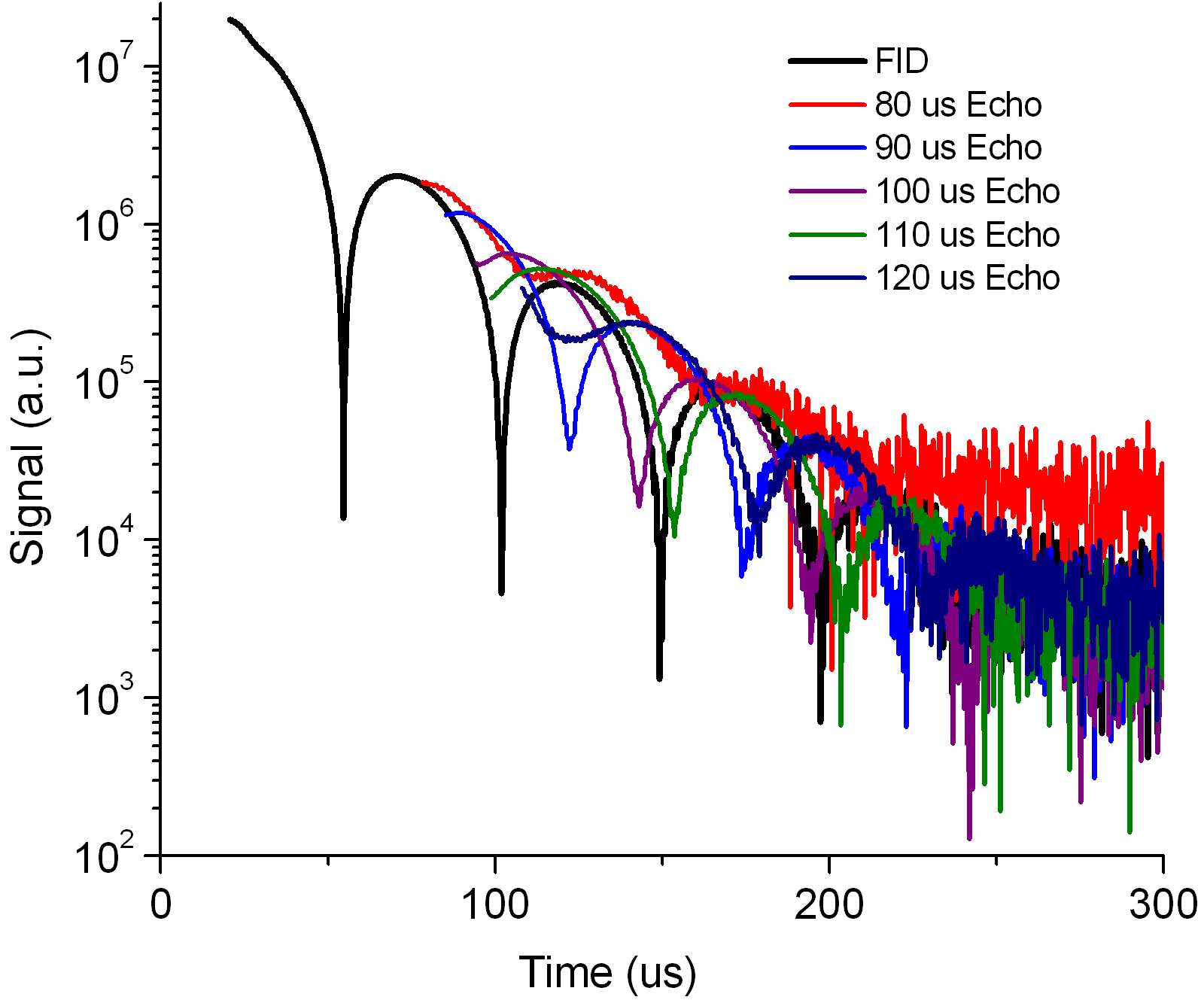}}\\
     \subfloat[]
  {\label{fig:c}\includegraphics[width=0.47\textwidth]{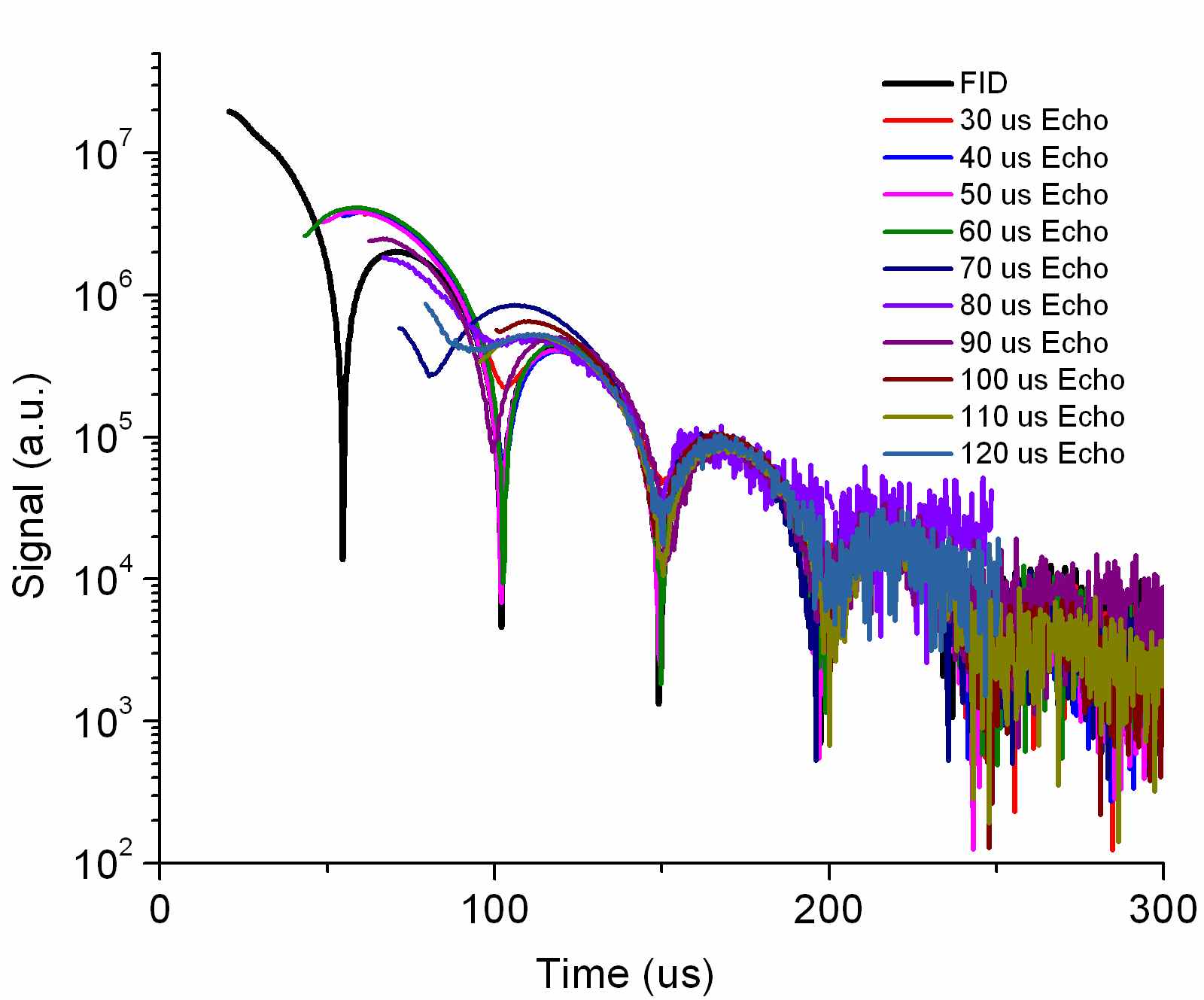}}   
  \caption{(Color online) $^{19}$F FID and solid echoes in single-crystal [111] CaF$_2$ (System IX; see Table~\ref{table4}).  In (a) and (b) we show ten solid echoes on a semilog plot together with the FID.  Signals are split between (a) and (b) for visual clarity only.  In (c), we show the same data as in (a) and (b) with the echoes time-shifted to illustrate the convergence of the long-time behavior.}
  \label{fig:caf2_111_se}
\end{figure*}

\begin{figure*}[htbp]
 % \centering
  \subfloat[]
  {\label{fig:powder_1}\includegraphics[width=0.47  \textwidth]{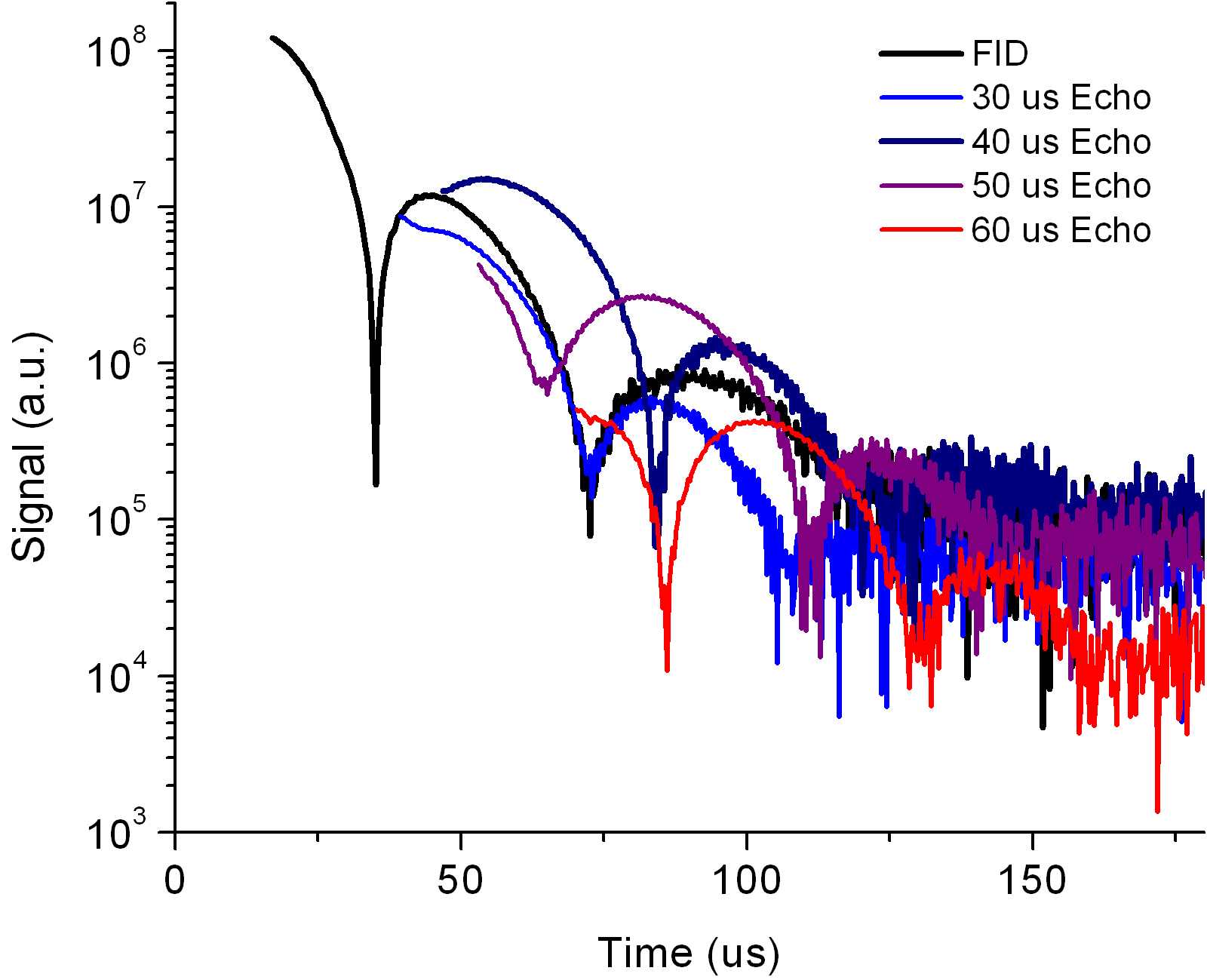}}                
\subfloat[]
  {\label{fig:powder_2}\includegraphics[width=0.47\textwidth]{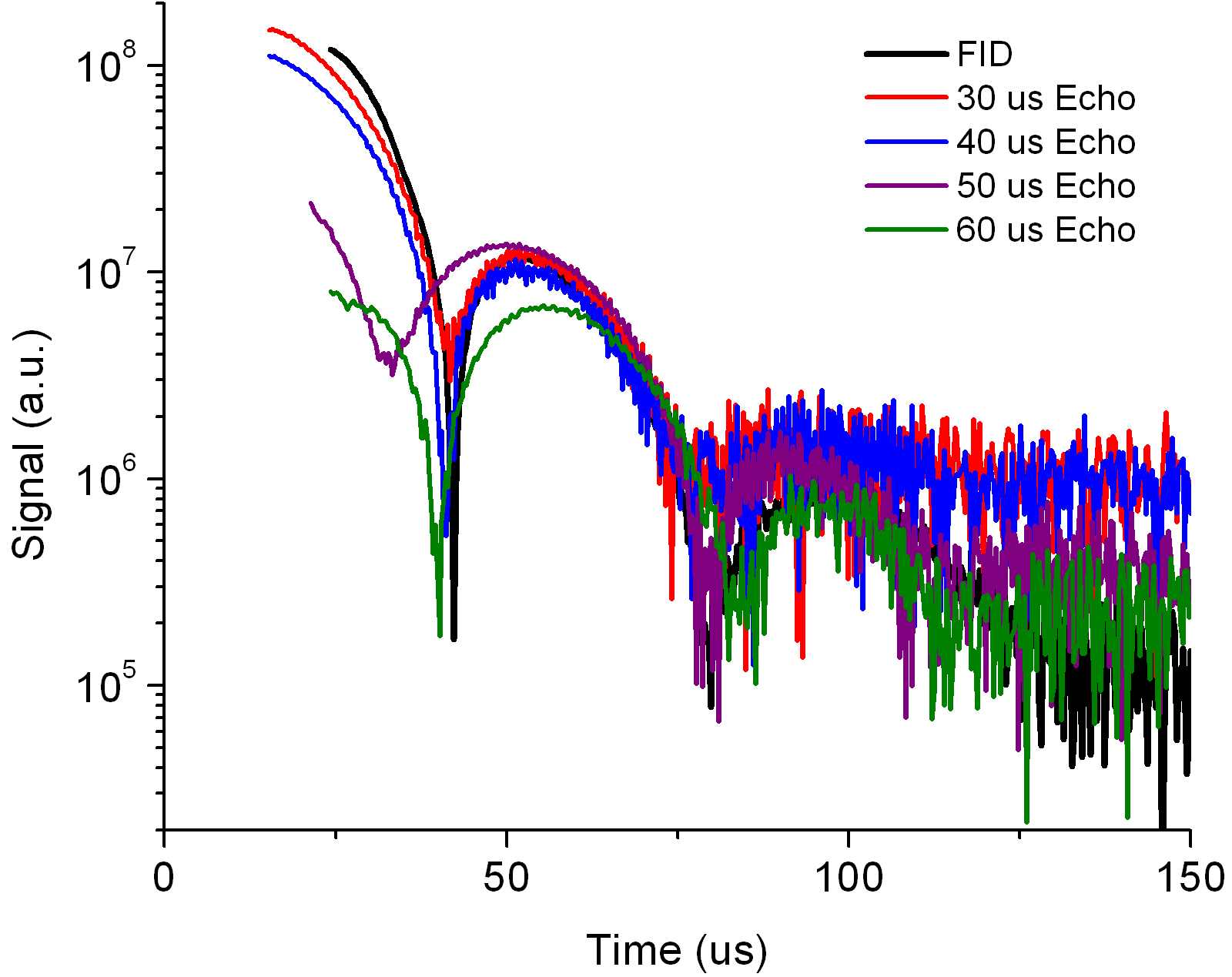}}
   %  \subfloat[]     
  \caption{(Color online) $^{19}$F signals in CaF$_2$ powder (System X; see Table~\ref{table4}).  In (a), solid echoes for four different values of $\tau$ are shown with the FID (black).  In (b), the solid echo signals are time- and amplitude-shifted to attempt to match the long-time portions of the signals.  The Lowe beats wash out after several times $T_2$, and they are neither similar to each other nor can they be characterized by a single beat frequency and decay coefficient within experimental resolution.}
  \label{fig:caf2_powder_se}
\end{figure*}

Fig.~\ref{fig:caf2_powder_se} shows the FID and four solid echo signals associated with the CaF$_2$ powder (System X; see Table~\ref{table4}).  The CaF$_2$ powder samples (as with all of the xenon samples) have many crystallites with random orientations; the transverse decays are therefore independent of sample orientation.  However, unlike in xenon, the initial beating seen in the signal washes out such that a single decay parameter and beat frequency are not present; therefore the long-time regime cannot be characterized by a universal behavior within our experimental resolution.  Furthermore, the time-shifts and amplitude manipulation of the solid echo signals shown in Fig.~\ref{fig:powder_2}  do not give conclusive evidence of a convergence of these signals to the behavior of the FID after several times $T_2$.  Polycrystalline CaF$_2$ was the only system we examined that, within our instrumental resolution, did not exhibit the universal long-time behavior predicted in Ref.\ 8.  
 
Solid xenon samples frozen from the liquid phase should also be in a polycrystalline state with randomly oriented crystalline regions, similar to a polycrystalline powder.  Previous work has shown~\cite{lowe_2}  (and we have verified here) that distinct orientations of single-crystal CaF$_2$ with respect to the external magnetic field produce different beat frequencies in the transverse decay.  The frequency differences are large enough that mixing them (as is the case in a polycrystalline powder) may cause the Lowe beating to wash out after a few times $T_2$ (see Fig.~\ref{fig:caf2_powder_se}).  One might have anticipated a similar wash out effect in polycrystalline xenon.  Instead, the solid xenon signals exhibit a very precise beat pattern that lasts over several orders of magnitude (see Fig.~\ref{fig:43A_FID}).  This phenomenon could be explained if the beat frequencies of a xenon single-crystal in different orientations with respect to the external magnetic field were similar or identical, or if the xenon were frozen in a glassy state with less order than a rigid polycrystalline structure.  Indeed, despite some claims in the early literature that it should be possible,~\cite{cohen} all recent attempts to vitrify monatomic liquids (noble gases, metals, or otherwise) have been unsuccessful to date despite achieving freezing rates on the order of $10^{10}$ K/s and pressures on the order of 10 GPa.  The only reported exception is a study which achieved vitrification of monatomic germanium by exploiting a particular property of tetrahedral liquids that does not exist in noble gases.\cite{vale}  No other monatomic liquid has yet been vitrified.  Our own X-ray diffraction data confirm that the frozen xenon is polycrystalline with no glassy component (see Fig.~\ref{fig:xray}).  

Why the xenon samples exhibit such well-defined beats and universal long-time behavior is an outstanding question; it is likely related to the difference between the fcc lattice of solid xenon and the simple cubic lattice formed by $^{19}$F in CaF$_2$.  Each lattice site has 12 nearest neighbors in the fcc lattice and 6 nearest neighbors in the simple cubic lattice, i.e., the arrangement of nearest neighbors is more spherically symmetric for the fcc lattice. As a result,  the powder average of the long-time beats should be less washed out in the fcc case. On the other hand, the CaF$_2$ powder may still exhibit well-defined beats at later times (not accessible in this experiment) once the crystal grains corresponding to the smallest value of gamma begin to dominate the overall signal.  We plan to address this question theoretically elsewhere.~\cite{unpub_1}  Unfortunately, noble-gas single-crystals can be generated only with considerable difficulty and in very small sizes.~\cite{allers} Furthermore, producing single-crystal hyperpolarized $^{129}$Xe would appear close to impossible, making direct experimental NMR measurements of single-crystal solid xenon prohibitively difficult. 

\subsubsection{\label{sec:magic_echo} The Magic Echo}

\begin{figure}[htbp]
\begin{center}
\includegraphics[width=0.47\textwidth]{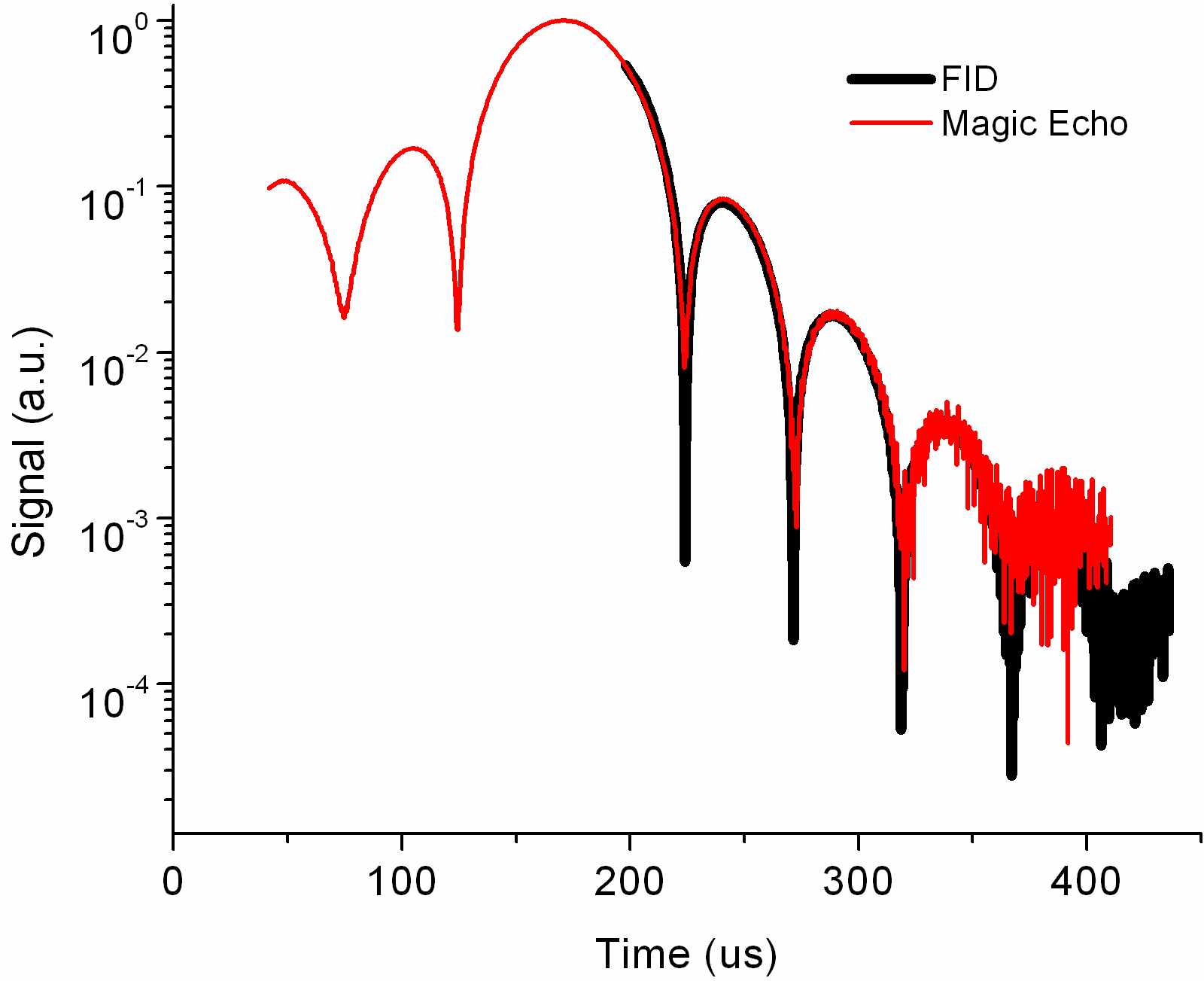}
\end{center}
\caption{(Color online) $^{19}$F FID and magic echo in single-crystal [111] CaF$_2$ (System IX; see Table~\ref{table4}).  Black line shows the FID, while the blue line maps out the signal obtained following the magic echo pulse sequence.  The time origin corresponds to the end of the final pulse in the magic echo sequence.  The magic echo signal begins after a $50\mu$s receiver recovery time.  The FID has been time-shifted to lie over the portion of the magic echo that reproduces the FID.  Note that the decays are indistinguishable even in the early (non-universal) portion of the FID signal.}
\label{fig:Magic_echo}
\end{figure}

To contrast our use of the solid echo and JB echo to generate distinct initial transverse spin configurations that then evolve to a universal long-time behavior, we used the magic echo~\cite{rhim} ($90_y - \tau - 90_{\bar{y}}  - \tau'_{H1} - \tau'_{\bar{H1}} - 90_y$) in [111] single-crystal CaF$_2$ as a form of control experiment to show that it is possible in these systems to reverse the sign of the Hamiltonian and reproduce the entire FID.  The resulting signal from the magic echo sequence shown in Fig.~\ref{fig:Magic_echo} is indeed found to have nearly the same shape as FID generated from a single $90^{\circ}$ pulse, notably even in the early portion of the FID decay.  For our experiment, $\tau = 11$ $\mu$s and $\tau'_{H_1} = \tau'_{\bar{H_1}} = 180$ $\mu$s.  The left-hand side of the magic echo near the time origin is not perfectly symmetric with the signal at $t>200\mu$s due to residual effects of the strong rf excitation pulse.  As expected,~\cite{slichter} the full dipolar refocusing (and therefore the maximum of the magic echo signal) occurs at a time $t = \tau'_{H_1} - \tau$. 

\section{\label{sec:discuss} Role of microscopic chaos}

It has been argued in Refs.~[29, 30] that the occurrence of the well-defined long-time behavior (1) in the absence of the separation of time scales is already a strong indication of the role of the microscopic chaos.  However, such a behavior could also appear in approximate theories of spin dynamics as a consequence of rather crude approximations. It was the prediction of the identical frequency and decay constants for the FID and all sorts of echoes \cite{Fine_1} that set apart the theory based on the notion of microscopic chaos from other theoretical descriptions, \cite{lowe_1, tijon, bork, bork_2, parker, engles_chao, lundin, jensen, fine_3} which, to the best of our knowledge, are not in position to make such a prediction. Below we elaborate on these issues.

Although the long-time behavior (1) is reminiscent of damped harmonic oscillator, one has to appreciate that damping in the latter case would originate from friction, which is usually caused by microscopic motion, which, in turn, is much faster than the period and the decay time of the damped oscillator. In the context of NMR, similar well-defined exponential damping appears in the problems of exchange narrowing and motional narrowing.~\cite{abragam, slichter}
 
It should be further noted that the occurrence of well-defined frequency beats  in the tail of FIDs and echoes is a somewhat counterintuitive effect, which is determined primarily by the negative relative sign of the $J^z$ and $J^{\perp}$ in Eq.\ (3). These beats, in particular, have little connection to the oscillations associated with a Pake doublet~\cite{pake, abragam} --- the latter originates from a highly discrete quantum structure of the FID problem for two isolated spins 1/2. In solid xenon and in CaF$_2$, each spin 1/2 is coupled to the whole lattice through many interacting neighbors. If the coupling constants $J^z$ and $J^{\perp}$ were of the same sign, the FIDs in CaF$_2$ and solid xenon would likely exhibit no beats, while the beats associated with a Pake doublet would remain with a modified frequency value.  The above emphasis on correlation effects is corroborated by simulations for a lattice of classical spins~\cite{BF-JSP} where similar beats have been observed --- clearly without any role for quantum discreteness. These and other factors determining the occurrence of beats are discussed in detail in Ref.\ 30.  Another difference between the Pake-doublet signal and the FIDs in CaF$_2$ or solid xenon is that the corresponding power spectrum in the former case has a well pronounced peak at the oscillation frequency, while in the latter case it exhibits no detectable signature of the long-time beats.~\cite{bruce}

In general, FID tails oscillate and decay on the fastest natural time scale of the nuclear spin system, which is the timescale of individual spin motion in the local field created by the neighbors. Historically, the absence of the separation of time scales was the reason why no universal shape for the FID tails was expected \emph{a priori}. Noteworthy in this respect was the proposal of Abragam to approximate FIDs in CaF$_2$ by function $e^{-a^2 t^2\over{2}} \sin({bt})/bt$, where the parameters $a$ and $b$ were to be determined from fitting of the exact values of the second and the fourth moments.~\cite{abragam}
  
The theoretical problem of calculating the FID in solids is non-perturbative.  Many attempts to deal with it resulted in various uncontrollable approximation schemes. These schemes aimed at reproducing the entire behavior of FID, of which the long-time tail was a rather insignificant part. At the same time, relatively crude approximations~\cite{tijon, bork_2, parker, engles_chao, lundin, jensen} made to predict the initial FID behavior sometimes resulted in the long-time behavior having form (1). 

A particularly strong claim of predicting the long-time behavior (1) was made by Borkmans and Walgraef.~\cite{bork_2} This claim, however, was criticized as inconclusive in Chapter 8 of Ref.~40, and in the appendix of Ref.~30.  Borkmans and Walgraef described the long-time behavior of FIDs by a memory-function-like integro-differential equation with memory function (integration kernel) having known initial behavior but unknown long-time behavior. Looking at the initial behavior, they assumed that the entire memory function had a Gaussian shape, and thereby obtained the long-time behavior of form (1). However, in fact,  the long-time behavior of the FID was entirely controlled by the unknown long-time behavior of the memory function: for the long-time FID behavior (1) to follow from the memory function equation, the asymptotic decay of the memory function should be no slower than exponential. If it were slower than exponential,  e.g. power law, then the asymptotic FID behavior would also exhibit a power law decay rather than (1).  

To the best of our knowledge, the theory of Borkmans and Walgraef has never been applied to solid echoes and other pulse sequences investigated in this work, and, in fact,  we are not sure whether, if applied, it would be manageable at all. If it were manageable, it is further difficult to see how that theory would reproduce the main result of this work, namely, the same long-time constants for different pulse sequences. Different pulse sequences would result in different initial conditions and hence  different parameters for the memory functions. It  then appears rather unlikely that  different memory functions would lead to the same long-time decay constants. A similar critique is applicable to all other approximations attempting to predict the long-time behavior of NMR signals from the knowledge of a few  initial time derivatives. 

The prediction of quantitatively identical long-time behavior for different pulse sequences was not made until the long-time behavior of NMR signals was considered on the basis of the notion of microscopic chaos.~\cite{Fine_1, BF-IJMPB}  The analysis of Ref.~30 has focused exclusively on the fundamental reasons behind the fact that the long-time functional form (1) is not a crude approximation but rather a robust accurate property. This analysis predicted that non-integrable lattices of interacting classical or quantum spins universally exhibit the long-time behavior (1), while the integrable cases amenable to the analytical calculations of FIDs exhibit non-universal behavior different from (1). Non-integrability in the interacting spin systems is the rule, while integrability is an exception. In classical spin systems, non-integrability is generally associated with microscopic chaos caused by non-linear interaction between spins. Mathematically, chaos is defined as an exponential instability of the phase space trajectories with respect to small differences of the initial conditions. Calculation of classical FIDs requires averaging  over all possible chaotic phase space trajectories. In the quantum case, the notion of chaos cannot be defined at the level of exponential instabilities in the phase space because the Heisenberg uncertainty principle precludes one from defining a phase space trajectory.  However, the connection between classical and quantum chaos can be conjectured for spin lattices at the level of spin correlation functions.~\cite{support, BF-IJMPB}

According to the analysis of Ref.~30, almost any initial probability distribution in a chaotic classical spin system would evolve to exhibit universal patterns associated with stable and unstable directions in the phase space. These patterns are complicated and intrinsic to a given Hamiltonian. They exhibit a dynamically developing small parameter representing the ratio of characteristic scales along stable and unstable phase space directions. This small parameter, in effect, substitutes for the separation of time scales and leads to Markovian long-time behavior of ensemble-averaged quantities on non-Markovian timescales [e.g.\ Eq.(1)]. As time progresses, the above patterns become increasingly singular and are expected to underlie chaotic eigenmodes of the time evolution operator. These eigenmodes are known as Pollicott-Ruelle resonances~\cite{ruelle, gaspard}. Their singular character is masked by ensemble averaging, but what remains after averaging are the exponential decays with or without oscillations. In terms of spin correlation functions, the formation of the above patterns imply the transfer of weight from lower-order to higher-order correlations (spin coherences in NMR language). It was conjectured in Refs.~[8, 20, 30] that the same transfer of spin correlations also occurs in quantum spins systems and underlies the observed exponential decays. The slowest of these decays then controls the long-time behavior in a given symmetry class of initial conditions. (NMR signals studied in this work correspond to the initial conditions having zero wave vector symmetry, i.e., all spins are equivalent from the viewpoint of initial conditions.) It was estimated in Ref.~30 that the onset of this single exponential behavior occurs in the NMR context on the scale of a few times $T_2$.  The prediction of the identical long-time behavior for different pulse sequences was based on the fact that the above Pollicott-Ruelle resonances and the associated exponential decays are intrinsic for a given Hamiltonian but not to the initial conditions, and therefore the exponential decay associated with the slowest resonance would dominate the behavior of many-spin density matrices and hence would manifest itself in the behavior of  all observable quantities in the same symmetry class independently of the initial condition.~\cite{Fine_1} 
  
At present, the theory of chaos in both classical and quantum many-body systems is still at an early development stage; the use of conjectures in the above analysis is therefore unavoidable. However, since the conjectures are not rigorously proven, it is essential to substantiate theoretical claims by extensive experimental tests, and this is what the present work does.  We also hope that the robustness of the long-time behavior (1) would stimulate the search for the theoretical methods of controllable calculations of parameters $\gamma$ and $\omega$ in Eq.(1). These methods, if found, are likely to have impact much beyond NMR theory.

\section{\label{sec:conclusion} Conclusion}

The present paper reports on a comprehensive investigation of the long-time behavior
of NMR signals well beyond the regimes explored in previous work, thereby establishing a new frontier of
what is known on this subject experimentally. In comparison with the previous study,~\cite{morgan} which investigated two polycrystalline samples of one material (solid xenon) in a smaller signal intensity range, the present study includes polycrystallites/powders and single crystals of two different materials (solid xenon and CaF$_2$) with a better signal resolution and with a much broader parameter space (multiple isotopic xenon concentrations, varied external fields for CaF2 crystals, and a wide variety of pulsed spin manipulations). In all cases except for the CaF$_2$ powder, we have observed sinusoidally modulated exponential long-time decays characterized by the same time constants irrespective of distinct initial conditions imposed by various spin manupulations.  In the CaF$_2$ powder, the sinusoidal Lowe beat pattern of $^{19}$F washes out within
our experimental resolution and therefore cannot be characterized by Eq.\ (1). In principle,
initial washing out of Lowe beats is expected in a powder where the beat frequencies of the various
crystallites are distinct.  In light of these complications associated with powder averaging, the fact that the well-defined universal behavior (1) is seen in the polycrystalline xenon systems is intriguing and needs to be better understood.  But more importantly, the results on single crystal CaF$_2$ first reported here emerge as more conclusive evidence of the identical long-time behavior of FIDs and various echoes. Since it is also not known \emph{a priori} how far the exponential-oscillatory long-time behavior can extend in time, particularly in powder samples, the improvement in the signal-to-noise ratio allowing one to explore longer time ranges is another essential advance of this study.

The bulk of this work provides evidence in support of the conjecture that a universal behavior will dominate the long-time transverse relaxation of manipulated spin coherences. This conjecture is based on the notion of microscopic chaotic mixing induced by the interaction between nuclear spins. We note that as our system for hyperpolarizing xenon nuclei continues to improve, the resulting highly spin-polarized lattice
will generate yet another class of unique transverse spin configurations that depart from the high-temperature limit;  these can then be examined for further verification of the universal long-time behavior.~\cite{ab_gold}  

\begin{acknowledgments}

We would like to thank M.\ P.\ Horvath for valuable help in acquiring the X-ray diffraction data as well as D.\ K.\ Solomon for providing the xenon gas assays.  We acknowledge very helpful discussions with M. S. Conradi, particularly in relation to JB echoes.  We also thank B. Meier and J. Haase, who are conducting similar CaF$_2$ experiments at Leipzig University, for sharing with us their preliminary data which appear to be consistent with our results.  This work was supported by NSF grant PHY-0855482.

\end{acknowledgments}

\newpage
\section{\label{sec:supp} Supplementary Section}
\begin{figure*}[h]
\noindent {\bf System II (62.7\% $^{129}$Xe)} \vspace{0.5in}
%FIGURE S1: System 2.
\renewcommand{\thefigure}{S1}
 % \centering
  \subfloat[]
  {\label{fig:145_a}\includegraphics[width=0.47  \textwidth]{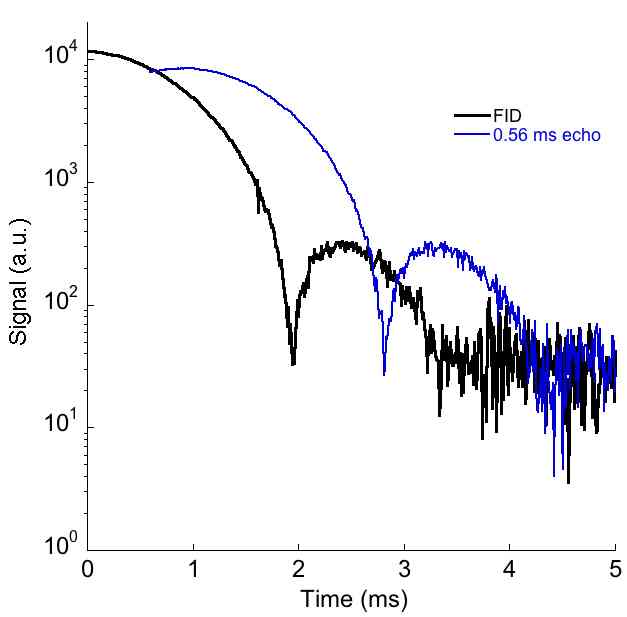}}                
\subfloat[]
  {\label{fig:145_b}\includegraphics[width=0.47\textwidth]{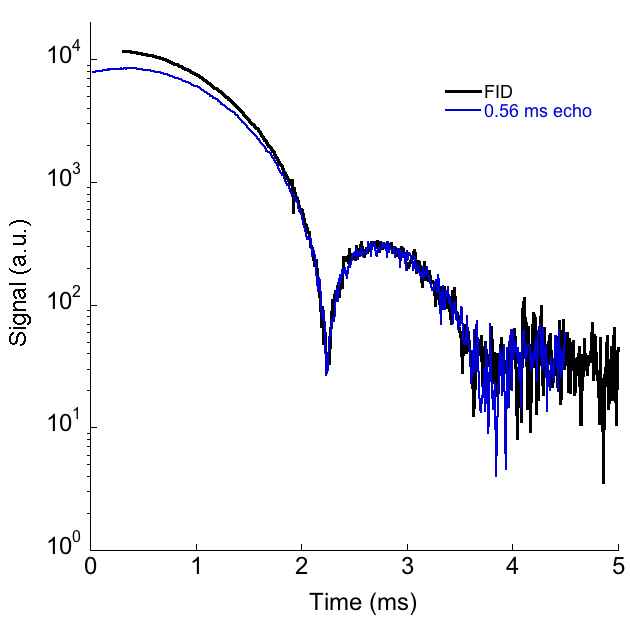}}
   %  \subfloat[]     
 \caption{(Color online) $^{129}$Xe free-induction decay and solid echo in xenon having $62.7\%$ $^{129}$Xe (System II; see Table I of main article).  (a) Solid echo signal shown on a semilog plot with the FID.  (b)  The same data as in (a) are shown again with the echo and FID time-shifted to illustrate the convergence of the long-time behavior.}
  \label{fig:145C}
\end{figure*}

%\clearpage

\begin{figure*}[h]
\noindent {\bf System III (54.7\% $^{129}$Xe)} \vspace{0.5in}
%FIGURE S2: System 3.
\renewcommand{\thefigure}{S2}
 % \centering
  \subfloat[]
  {\label{fig:145_a}\includegraphics[width=0.5  \textwidth]{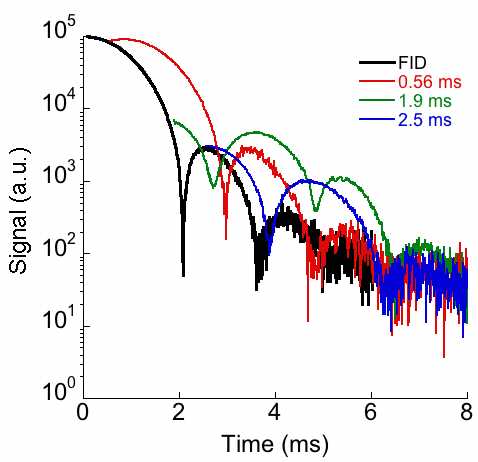}}                
\subfloat[]
  {\label{fig:145_b}\includegraphics[width=0.5\textwidth]{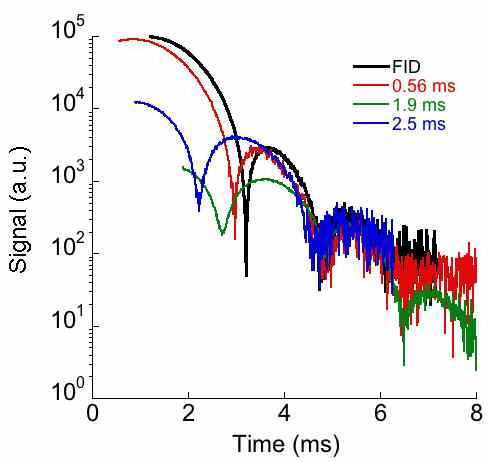}}
   %  \subfloat[]     
 \caption{(Color online) $^{129}$Xe free-induction decay and solid echoes in xenon having $54.7\%$ $^{129}$Xe (System III; see Table I of main article).  (a) Solid echo signals shown on a semilog plot with the FID.  (b)  The same data as in (a) are shown again with the echoes time-shifted to illustrate the convergence of the long-time behavior.}
  \label{fig:146A}
\end{figure*}
\clearpage

\begin{figure*}[h]
\noindent {\bf System IV (46.5\% $^{129}$Xe)} \vspace{0.5in}
%FIGURE S3: System 4.
\renewcommand{\thefigure}{S3}
% \centering
  \subfloat[]
  {\label{fig:145_a}\includegraphics[width=0.5  \textwidth]{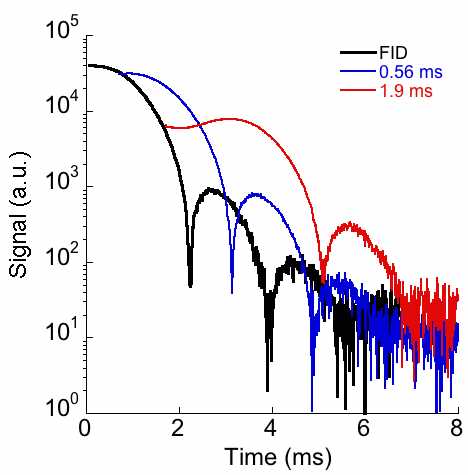}}                
\subfloat[]
  {\label{fig:145_b}\includegraphics[width=0.5\textwidth]{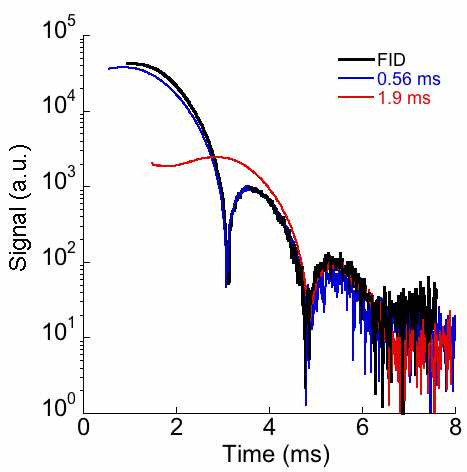}}
   %  \subfloat[]     
 \caption{(Color online) $^{129}$Xe free-induction decay and solid echoes in xenon having $54.7\%$ $^{129}$Xe (System IV; see Table I of main article).  (a) Solid echo signals shown on a semilog plot with the FID.  (b)  The same data as in (a) are shown again with the echoes time-shifted to illustrate the convergence of the long-time behavior.}
  \label{fig:146B}
\end{figure*}

%\clearpage
\begin{figure*}[h]
\noindent {\bf System VI (27.5\% $^{129}$Xe)} \vspace{0.5in}
%FIGURE S5: System 6
\renewcommand{\thefigure}{S4}

 % \centering
  \subfloat[]
  {\label{fig:1}\includegraphics[width=0.5 \textwidth]{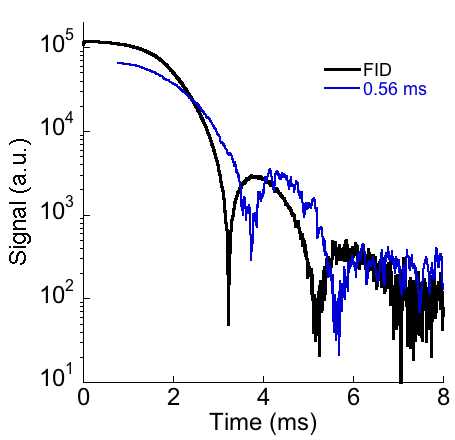}}                
\subfloat[]
  {\label{fig:2}\includegraphics[width=0.5\textwidth]{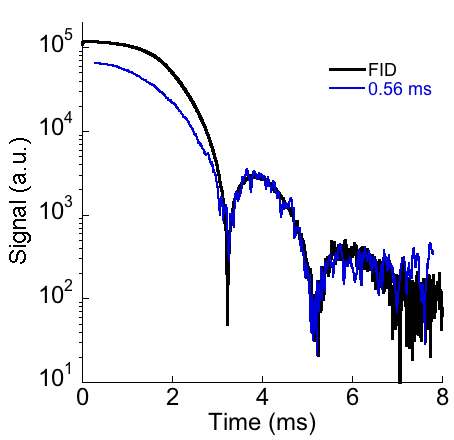}} 
  \caption{(Color online) $^{129}$Xe free-induction decay and solid echoes in xenon having $27.5\%$ $^{129}$Xe (System VI; see Table I of main article).  (a) Solid echo signal shown on a semilog plot with the FID.  (b)  The same data as in (a) are shown again with the echoes time-shifted to illustrate the convergence of the long-time behavior.}
  \label{fig:123A_se}
\end{figure*}

%\clearpage
\begin{figure*}[h]
\noindent {\bf System VII ([100] CaF$_2$ crystal)} \vspace{0.5in}
%FIGURE S6: System 7
\renewcommand{\thefigure}{S5}

 % \centering
  \subfloat[]
  {\label{fig:1}\includegraphics[width=0.45  \textwidth]{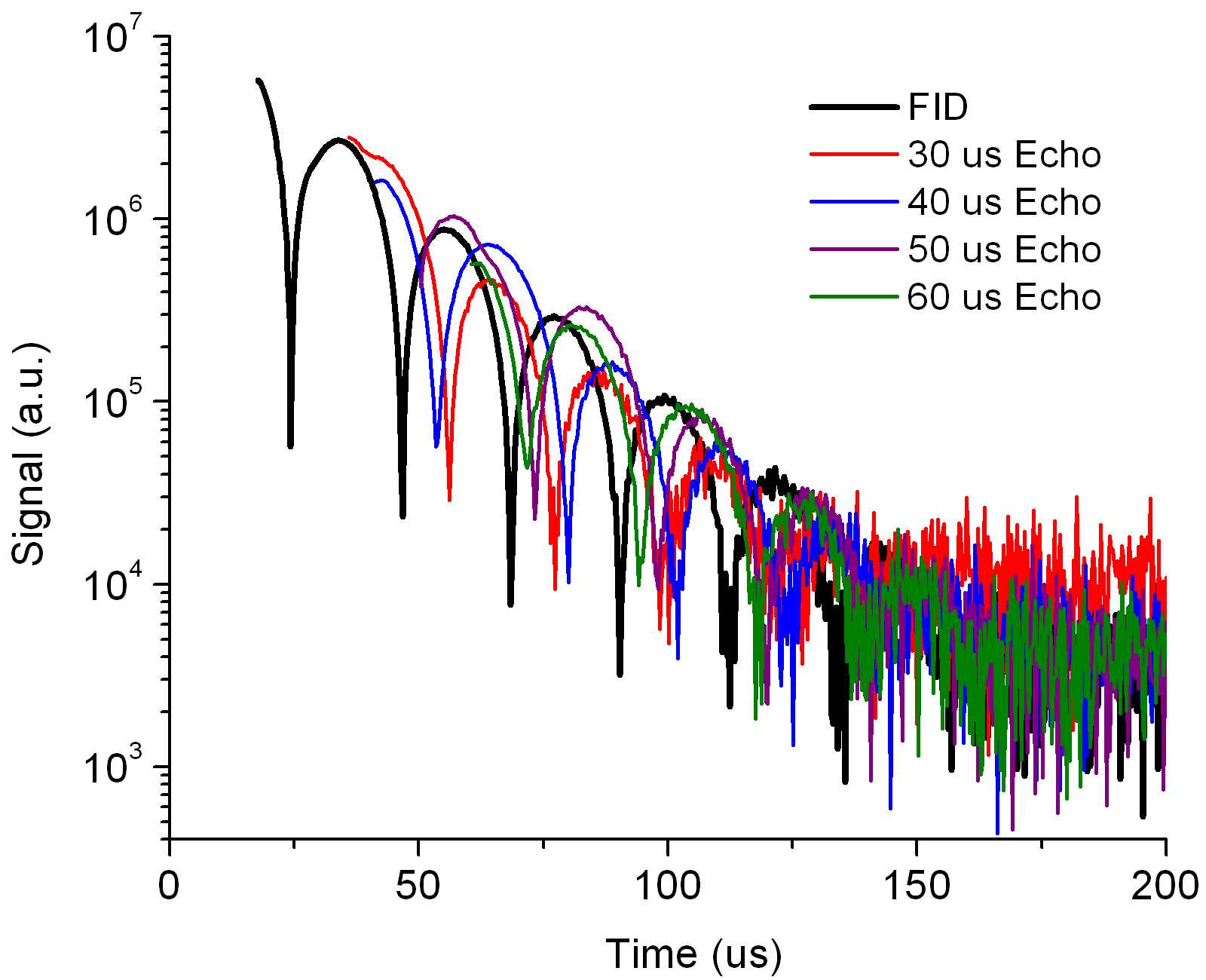}}                
\subfloat[]
  {\label{fig:2}\includegraphics[width=0.45\textwidth]{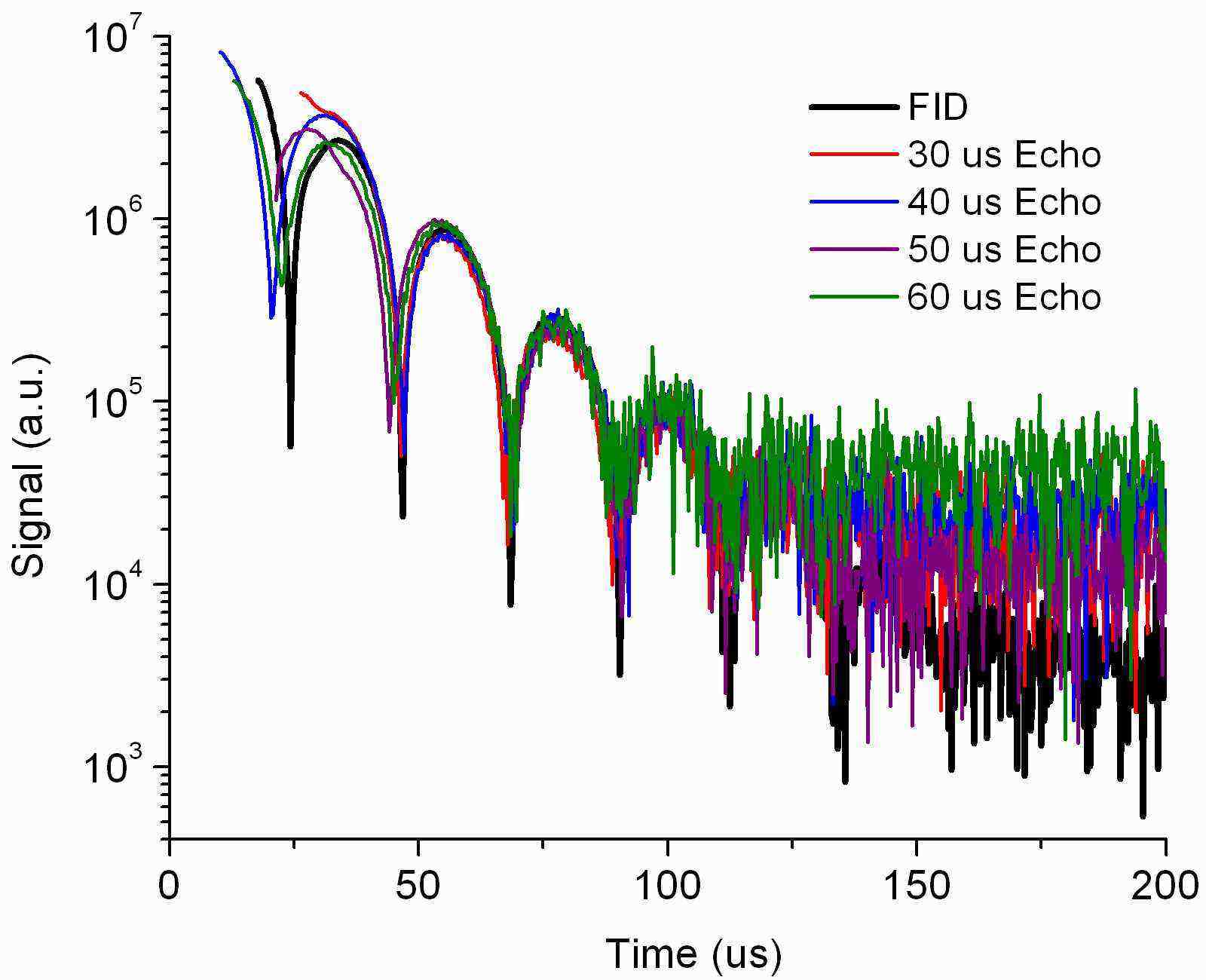}} \\
 \subfloat[]
 {\label{fig:3}\includegraphics[width=0.45\textwidth]{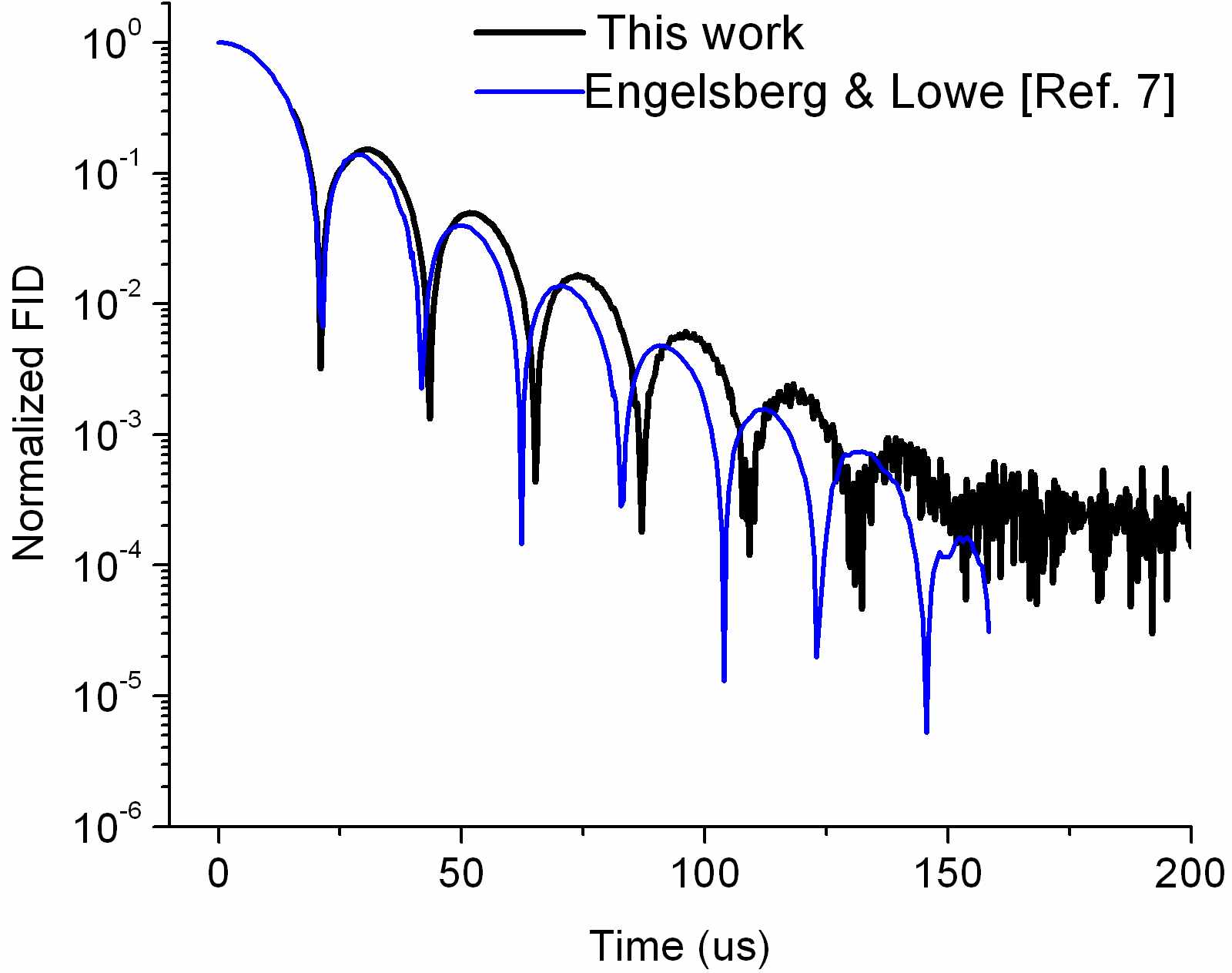}} 
  \caption{$^{19}$F FID and solid echoes in single-crystal CaF$_2$ with external field along [100] (System VII; see Table III of main article).  (a) Solid echo signals shown on a semilog plot with the FID.  (b) The same data as in (a) are shown again with the echoes time-shifted to illustrate the convergence of the long-time behavior.  (c) FID obtained in this work plotted with that obtained in Ref.~7.  Discrepancies in the beat frequency can be attributed to slight differences in crystal alignment.}
  \label{fig:caf2_100_se}
\end{figure*}

%\clearpage
\begin{figure*}[h]
\noindent {\bf System VIII ([110] CaF$_2$ crystal)} \vspace{0.5in}
%FIGURE S7: System 8
\renewcommand{\thefigure}{S6}

 % \centering
  \subfloat[]
  {\label{fig:1}\includegraphics[width=0.45  \textwidth]{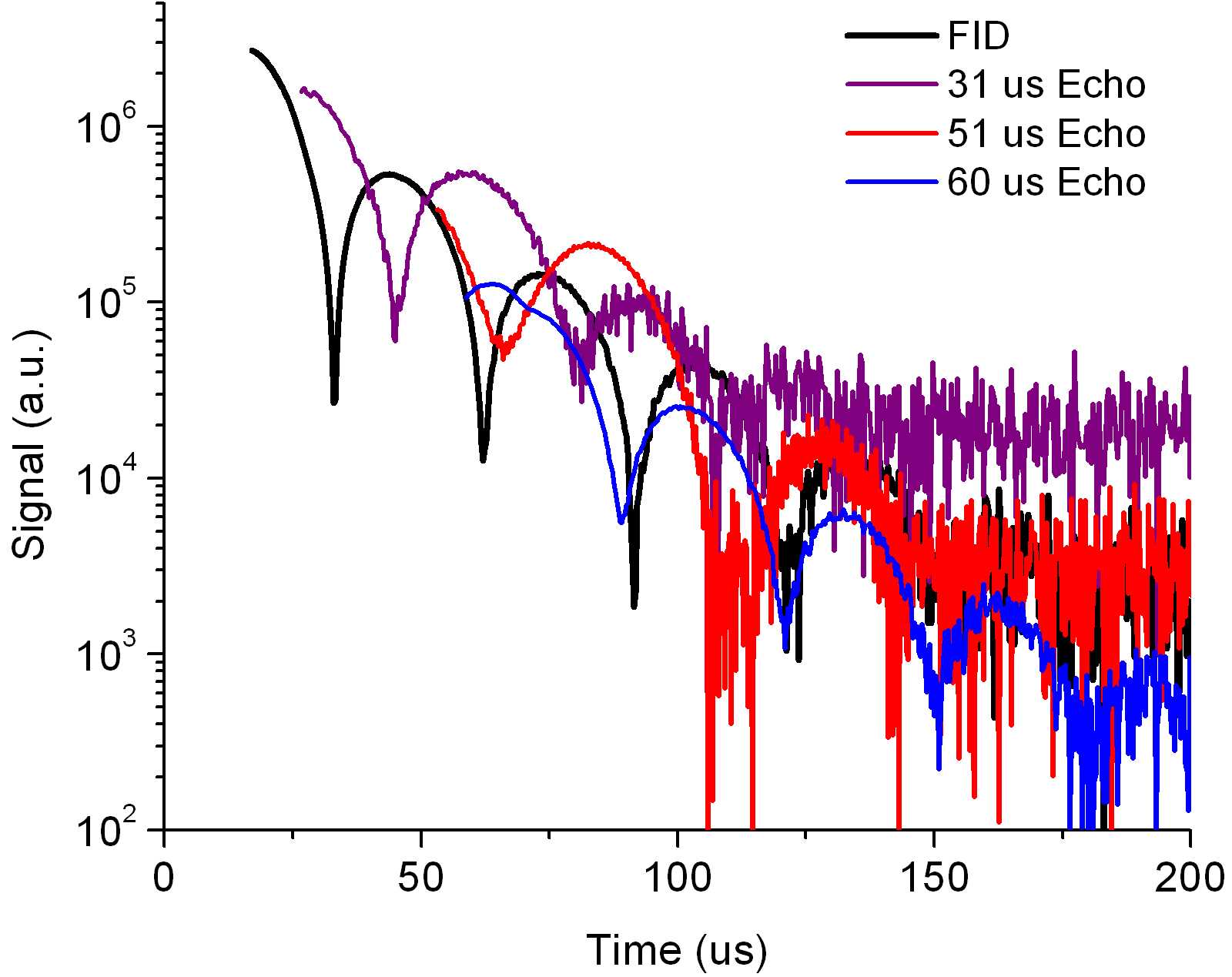}}                
\subfloat[]
  {\label{fig:2}\includegraphics[width=0.45\textwidth]{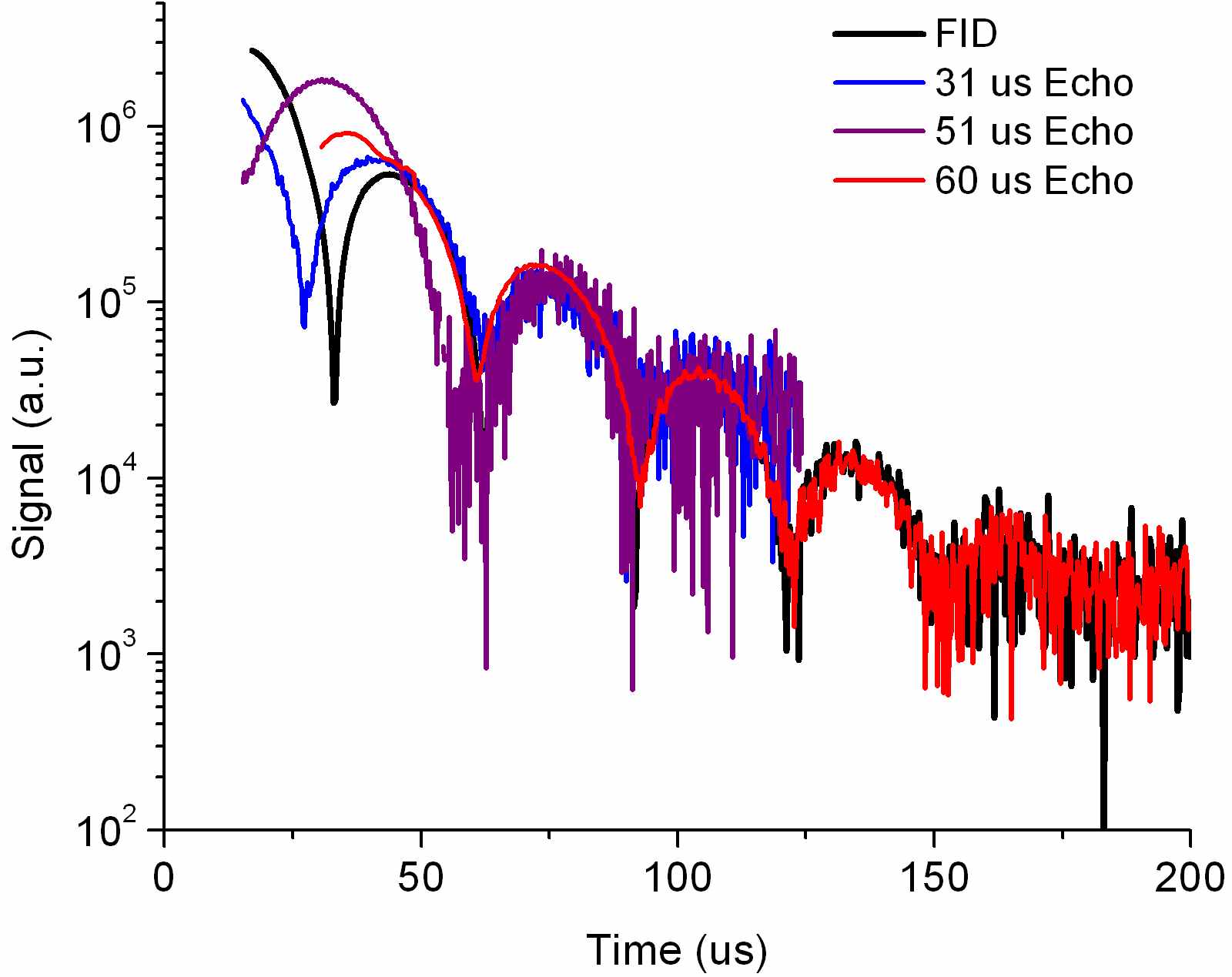}} \\
 \subfloat[]
 {\label{fig:3}\includegraphics[width=0.45\textwidth]{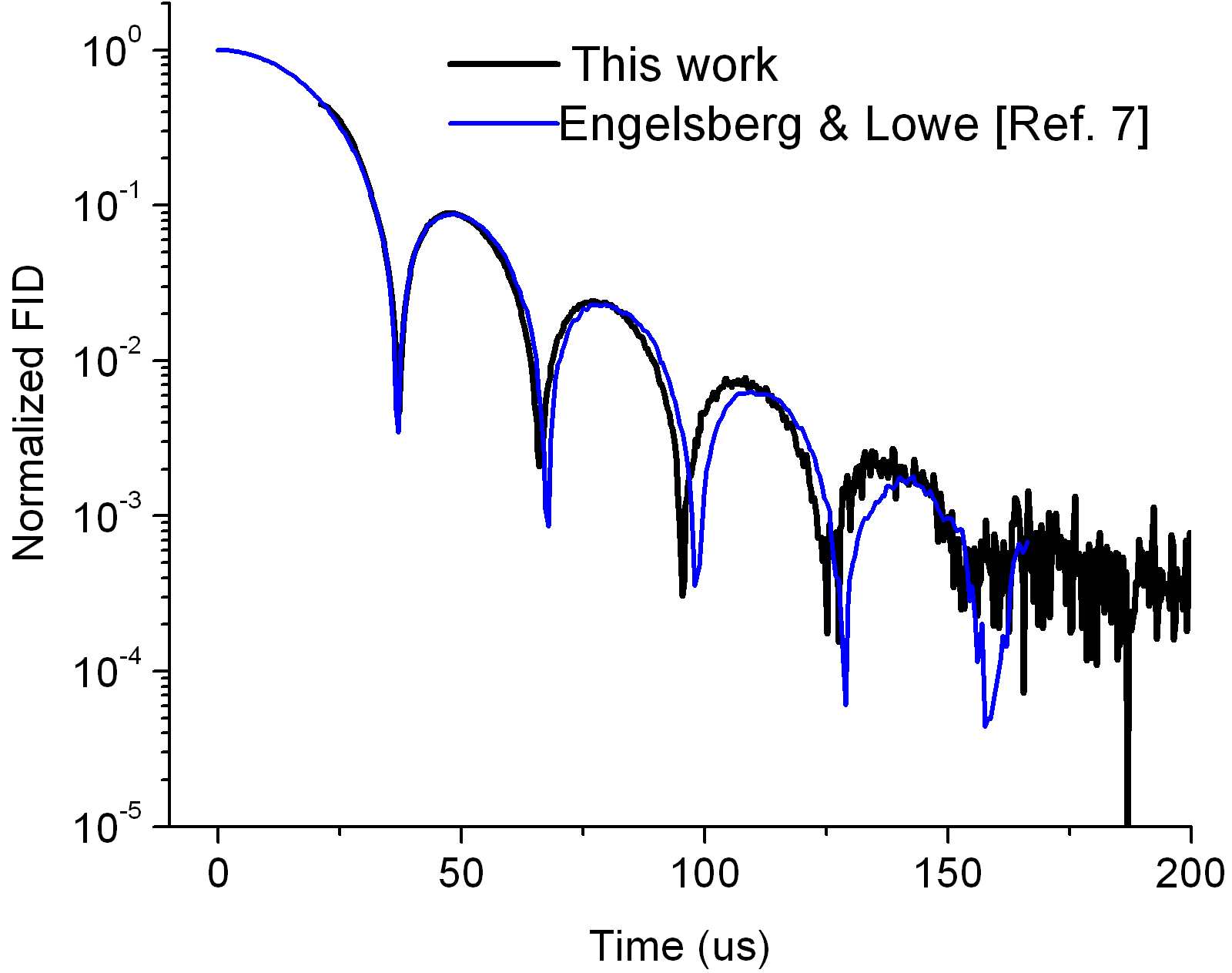}}
   %  \subfloat[]     
  \caption{$^{19}$F FID and solid echoes in single-crystal CaF$_2$ with external field along [110] (System VIII; see Table III of main article).  (a) Solid echo signals shown on a semilog plot with the FID.  (b) The same data as in (a) are shown again with the echoes time-shifted to illustrate the convergence of the long-time behavior.  (c) FID obtained in this work plotted with that obtained in Ref.~7.  Discrepancies in the beat frequency can be attributed to slight differences in crystal alignment.}
  \label{fig:caf2_110_se}
\end{figure*}

%\clearpage
\begin{figure*}[htbp]
\noindent {\bf System IX ([111] CaF$_2$ crystal)} \vspace{0.5in}
%FIGURE S8: System 9
\renewcommand{\thefigure}{S8}

 \subfloat[]
  {\label{fig:a}\includegraphics[width=0.44\textwidth]{caf2_111_early_unshifted.jpg}}                
\subfloat[]
  {\label{fig:b}\includegraphics[width=0.44\textwidth]{caf2_111_late_unshifted.jpg}}\\
     \subfloat[]
  {\label{fig:c}\includegraphics[width=0.44\textwidth]{caf2_111_shifted.jpg}}
 \subfloat[]
 {\label{fig:3}\includegraphics[width=0.44\textwidth]{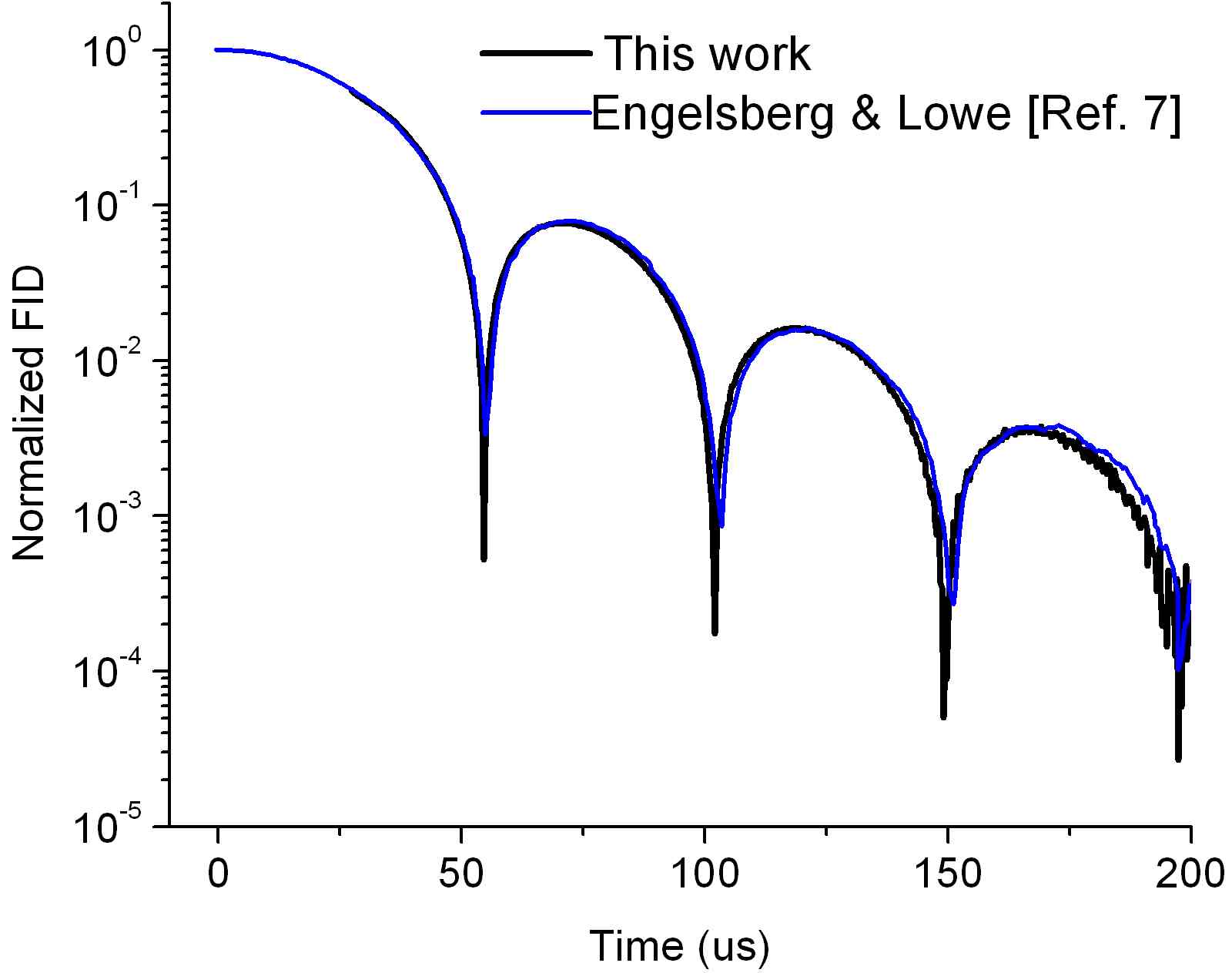}}   
  \caption{(Color online) $^{19}$F FID and solid echoes in single-crystal CaF$_2$ with external field along [111] (reproduced from Fig. 7 in main atricle).  In (a) and (b) we show ten solid echoes on a semilog plot together with the FID.  Signals are split between (a) and (b) for visual clarity only.  In (c), we show the same data as in (a) and (b) on a semilog plot with the echoes time-shifted to illustrate the convergence of the long-time behavior.  (d) FID obtained in this work plotted with that obtained in Ref.~7.  Discrepancies in the beat frequency can be attributed to slight differences in crystal alignment.}
  \label{fig:caf2_111_se}
\end{figure*}

\end{document}